\documentclass[journal]{vgtc}                % final (journal style)
\ifpdf%                                % if we use pdflatex
  \pdfoutput=1\relax                   % create PDFs from pdfLaTeX
  \usepackage{graphicx}                % allow us to embed graphics files
  \DeclareGraphicsExtensions{.pdf,.png,.jpg,.jpeg} % for pdflatex we expect .pdf, .png, or .jpg files
\else%                                 % else we use pure latex
  \usepackage{graphicx}                % allow us to embed graphics files
  \DeclareGraphicsExtensions{.eps}     % for pure latex we expect eps files
\fi%

\graphicspath{{figures/}{pictures/}{images/}{./}} % where to search for the images

\usepackage{microtype}                 % use micro-typography (slightly more compact, better to read)
\PassOptionsToPackage{warn}{textcomp}  % to address font issues with \textrightarrow
\usepackage{textcomp}                  % use better special symbols
\usepackage{mathptmx}                  % use matching math font
\usepackage{times}                     % we use Times as the main font
\usepackage{cite}                      % needed to automatically sort the references
\usepackage{tabu}                      % only used for the table example
\usepackage{booktabs}                  % only used for the table example
\usepackage{balance}
\usepackage{soul} %allow highlighting text
\usepackage{xcolor,colortbl} % for coloring table xxx
\usepackage{multirow}

\usepackage{amssymb} 

\def\revcolor{black} %_b
\newcommand{\rev}[1]{\textcolor{\revcolor}{#1}}

\preprinttext{
}

\onlineid{1062}

\vgtccategory{Research}
\vgtcpapertype{evaluation}

\title{VocabulARy: Learning Vocabulary in AR Supported by Keyword Visualisations}

\author{
Maheshya Weerasinghe\thanks{University of Primorska, Koper, Slovenia},
Verena Biener\thanks{Coburg University of Applied Sciences, Germany}, 
Jens Grubert\thanks{Coburg University of Applied Sciences, Germany}, 
Aaron J Quigley\thanks{Computer Science and Engineering , UNSW, Sydney, NSW, Australia},
\\\
Dr Alice Toniolo\thanks{School of Computer Science, University of St Andrews, St Andrews, UK}, 
Klen Čopič Pucihar \thanks{University of Primorska, Koper, Slovenia}, %\secthanks{Faculty of information studies, Novo Meesto, Slovenia}, 
Matjaž Kljun\thanks{University of Primorska, Koper, Slovenia}, 
}
\authorfooter{
}

\abstract{Learning vocabulary in a primary or secondary language is enhanced when we encounter words in context. This context can be afforded by the place or activity we are engaged with. Existing learning environments include formal learning, mnemonics, flashcards, use of a dictionary or thesaurus, all leading to practice with new words in context. In this work, we propose an enhancement to the language learning process by providing the user with words and learning tools in context, with VocabulARy. VocabulARy visually annotates objects in AR, in the user’s surroundings, with the corresponding English (first language) and Japanese (second language) words to enhance the language learning process. In addition to the written and audio description of each word, we also present the user with a keyword and its visualisation to enhance memory retention. We evaluate our prototype by comparing it to an alternate AR system that does not show an additional visualisation of the keyword, and, also, we compare it to two non-AR systems on a tablet, one with and one without visualising the keyword. Our results indicate that AR outperforms the tablet system regarding immediate recall, mental effort and  task-completion time. Additionally, the visualisation approach scored significantly higher than showing only the written keyword with respect to immediate and delayed recall and learning efficiency, mental effort and task-completion time.     
} % end of abstract March 9th 2022


\teaser{
  \centering
  \includegraphics[width=\linewidth]{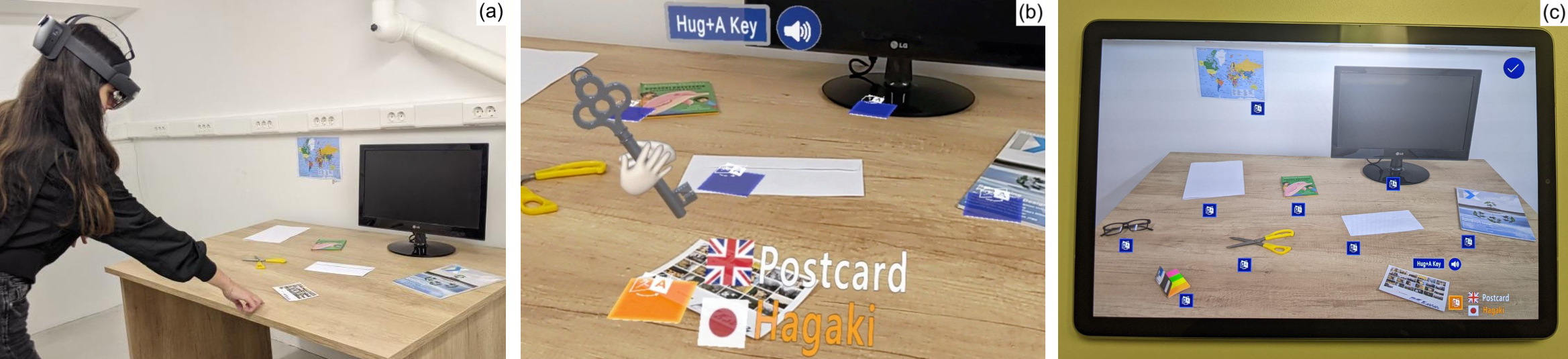}
  \caption{
   VocabulARy prototype. (a) Participant interacting with VocabulARy during the study; (b) VocabulARy prototype though HoloLens2 in \textsc{keyword + visualisation} instruction mode (the Japanese word ``hagaki'' sounds as the English phrase ``hug a key'' (keyword) and visualised with an animated hand grabbing a key (visualisation)); (c) Participant interacting with non-AR version of VocabulARy in \textsc{keyword} instruction mode (Note there is no visualisation of the keyword). AR and non-AR condition were tested with both instruction modes.  %\hl{Aaron says: I suggest 2 or 3 images from different stages of the same scenario showing very different aspects of VocabulARy}
  }
  \label{fig:teaser}
}

\vgtcinsertpkg

\begin{document}



\firstsection{Introduction}
\maketitle
%% \section{Introduction} %for journal use above \firstsection{..} instead
%% NEW (9th March 2022 edited) 
Learning a language is a complex task that requires dedication, perseverance and hard work. The basic learning process consists of comprehension of input (i.e.\ hearing or reading), comprehensible output (i.e.\ speaking or writing) and feedback (i.e.\ identifying errors and making changes in response) \cite{nunan1999second, brown1995readings}. Through these processes we learn vocabulary and grammar enhancing our language comprehension and expression abilities.  

Expanding one's vocabulary is an essential element of language learning and in vocabulary learning, methods for improving learners' memory play a vital role. Mnemonics is one such effective method, in which the learner attempts to link new learning with prior knowledge through the use of visual and/or acoustic cues. Keywords are one such practical technique in which the learner attempts to create a symbolic link between new and prior knowledge using associations triggered by keywords, a method shown to be particularly effective in prior research~\cite{atkinson1975mnemotechnics}.  

Furthermore, previous research shows that learning vocabulary can be enhanced through an encounter with words in context~\cite{pressley1982}. 
Existing learning environments include formal learning,
flashcards, use of a dictionary or thesaurus, all leading to practice with new words in context. For example, in formal learning the context is provided by the instructor or the provided instructional materials, 
%in mnemonics the context is provided by trying to link new learning with prior knowledge through the use of visual and/or acoustic cues,
%which is know as an effective strategy to boost students memory. 
in flashcards it is formed through images depicted on physical cards, in thesaurus it is provided through the provision of synonyms.

Consumer devices such as smartphones, tablets and head mounted displays can be used to enhance existing learning environments or to provide new ones. These systems enable technology driven paradigm shifts such as e-learning~\cite{Nicholson2007, mayes2007learning,mcloughlin2010personalised}, and more recently m-learning (mobile learning)~\cite{kumar2018learning, georgiev2004m, Spitzer2015}. All are capable of enhancement through better provision of learning context and methods for improving learners' memory. Furthermore, these systems are also capable of running Augmented Reality (AR) applications which have the potential to make language learning more intuitive and immersive because of their intrinsic ability to visualise digital information within a real world context. This is particularly important for vocabulary learning because it allows word encounters in real-world context, an important catalyst for vocabulary learning~\cite{santos2016augmented,vazquez2017serendipitous,ibrahim2018arbis}.

Despite the fact that prior work looked at AR for vocabulary learning discovering several benefits, such as better improved retention, higher enjoyment, motivation and engagement, 
none provide a direct comparison of AR \rev{applications that run in head-mounted displays} to the same technique within a non-AR interface.  Furthermore, to the best of our knowledge, no existing evaluation of vocabulary learning that combines keywords with visualisations exist. 

This paper contributes to addressing this gap with VocabulARy, an AR application for vocabulary learning that visually annotates objects in AR, in the user’s surroundings, with the corresponding English (first language) and Japanese (second language) words. In addition to the written and audio description of each word, VocabulARy also presents the user with a keyword and its visualisation to enhance memory retention. 
We evaluate the VocabulARy prototype by comparing it to an alternate
AR system that does not show an additional visualisation of the keyword and also, we compare it to two non-AR systems on a tablet,
one with and one without visualising the keyword. The results show that \textsc{ar} outperforms the \textsc{non-ar} (tablet) system regarding short-term retention, mental effort and  task-completion time. Additionally, the visualisation approach scored significantly higher than only showing the written keyword with respect to immediate and delayed recall and learning efficiency, mental effort and task-completion time.

%% OLD (8th March 2022 edited) 
% \begin{itemize}
%     \item AR has potential to make language learning more intuitive and immersive
    
%     \item multiple previous work that has looked at AR for vocabulary learning but never compared to "the same" non-AR technique
    
%     \item previous work suggest better long-term retention

%     \item combining AR with keyword-method

%     \item keyword method has found to be efficient, especially for high imagery words
    
%     \item our contributions are: 1) compare AR to a very similar non-AR system to see advantages of AR without confounding factors like learning-strategy 2) combine AR vocabulary learning with the keyword method 3) evaluate if an additional visualisation of the keyword improves vocabulary learning
% \end{itemize}

%%%%%%%%%%%%%%%%%%%%%%%%%%%%%%%%%%%%%%%%%%%%%%%%%%%%%%%%%%%%%%%%%%%%%%%%%%%%%%%%%%%%%%%%%%%%%%%%%%%%%%
%%%%%%%%%%%%%%%%%%%%%%%%%%%%%%%%%%%%%%%%%%%%%%%%%%%%%%%%%%%%%%%%%%%%%%%%%%%%%%%%%%%%%%%%%%%%%%%%%%%%%%
\section{Related Work}
\label{sec:relatedWork}
% Our work combines two areas. First, language learning in Augmented Reality (AR), specifically for learning vocabulary. Second, provide instruction mode for vocabulary learning, particularly using the keyword method.

% Learning is accomplished primarily through the process of comprehending learning materials in a self directed way~\cite{lee2009}. Presentations of information in learning materials often include several modalities, scuh as  text, audio and/or visuals. Compared to other methods visuals 
% %used with instructional support 
% offer several potential benefits for learning. For example, previous research showed recall and memory are better when information is presented though visuals instead of text.~\cite{clark2010,standing1970,lohr2007}. Besides, visuals can also arouse learners’ interest, curiosity, and motivation~\cite{mayer1998,mayer2007}. 

Vocabulary learning  can be enhanced through methods for improving learners’ memory~\cite{putnam2015,mastropieri1991} or through an encounter with words in context~\cite{pressley1982}. AR is an emerging technology for learning in real-world context and to scaffold this we structure our related work into: Learning context, Vocabulary learning in AR and Memory enhancement techniques. To better position our work in the context of language learning in AR environments we also classify prior work based on AR devices, learning content, presentation and learning method (\autoref{tab:relatedwork}).

\begin{table}[th]
  \caption{Selected prior work related to language learning in AR environments.}
  \centering
  \begin{tabular}{c}
      \includegraphics[width=0.98\columnwidth]{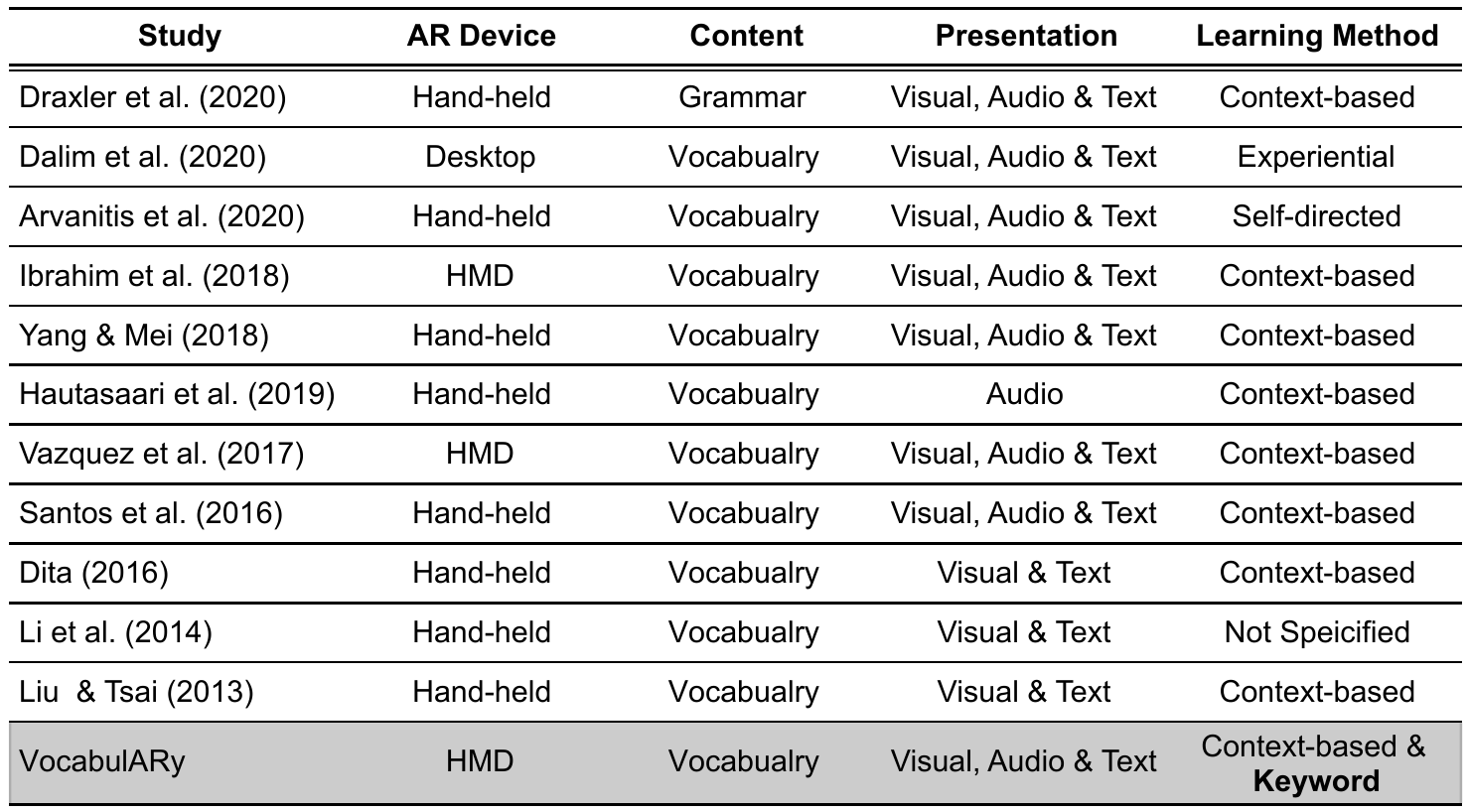} 
  \end{tabular}
  \label{tab:relatedwork}
\end{table}

\subsection{Learning Context}
It has been shown that people are more motivated to learn, if they can see the importance of the content with respect to the situation or, if they find the content interesting~\cite{pintrich2003motivational}. For example, being in a bar in a foreign country is likely to increase the interest in learning words and sentences required for ordering a coffee. Additionally, the context makes it possible to form associations that help later retrieval in similar circumstances~\cite{yang2018understanding,ibrahim2018arbis,vazquez2017serendipitous,santos2016augmented}. In other words, new words relevant to the learning context are more likely to be recalled than unrelated words~\cite{edge2011micromandarin}. 

AR has the ability to provide context-specific information in an interactive manner. In addition, AR can take any situation, location, environment, or experience to a whole new level by combining digital information with real-world contents. Thus, it has the potential to create more engaging and immersive learning environments. There exists a considerable body of previous work on AR systems that support learning in real-world contexts. For instance, there are systems that provide labels of new words corresponding to real-world objects~\cite{santos2016augmented,draxler2020augmented}, while others create imaginary settings to describe and enhance the physical properties of everyday objects~\cite{ibanez2014experimenting,strzys2018physics}.

\subsection{Vocabulary Learning in AR}
\label{sec:relatedWorkAR}
Previous studies have shown that AR offers many advantages for language and vocabulary learning. For instance, some studies reported that AR improved learning achievements and boosted motivation, engagement and collaboration among learners~\cite{hautasaari2019vocabura,santos2016augmented,ibrahim2018arbis,dalim2020using,draxler2020augmented}. \rev{Despite, some technical limitations of using AR for learning should be taken into account such as such as educators' limited proficiency with the relatively new technology \cite{khoshnevisan2021affordances} or the trade-off between connecting the experience to the context of the current location and providing a flexible and portable experience \cite{wu2013current}.}

Fujimoto et al.~\cite{fujimoto2012relation} \rev{have shown that users can memorise AR information better if it is shown within the location of a target object in the real world (e.g. AR information about a country shown over a map within this country). However, the information to be memorised in study did not take the context of the real environment and users' surroundings into account.}
%However, their study is limited by the fact that information outside the real world context is used.

% Godwin-Jones et al. \cite{godwin2016augmented} give an overview of AR technologies for language learning. 
% And since then, there have been different approaches to using AR to for vocabulary learning.

Several studies presented applications for learning vocabulary using hand-held AR devices. Hautasaari et al.~\cite{hautasaari2019vocabura} developed the VocaBura smartphone application for learning vocabulary during dead time. The application tracks a users' GPS locations and presents vocabulary related to the current location via audio. A study comparing this to an audio-only method showed that 7 days after the study, participants could recall significantly more words. Santos et al.~\cite{santos2016augmented} presented a handheld AR system that displays text, images, animation and sound next to corresponding real-world objects to learn Filipino or German. They compared this system to a non-AR tablet application using a flash card method.
Their results indicate that for tests directly after the experiment non-AR users performed better, yet this difference was not detected for long-term retention.
%use fiducial markers to recognize objects

Positive effects of AR technology in the context of increased motivation and enjoyment have also been detected. For example, Dalim et al.~\cite{dalim2020using} presented a system combining desktop-AR and speech recognition (TeachAR) and found that it increases children's knowledge gain and enjoyment. Similarly, Li et al.~\cite{li2014pilot} explored an AR application for language learning and found that it increased motivation in the beginning, yet for most participants motivation decreased at the end of the study.

The existing body of literature also includes applications for learning vocabulary on AR head-mounted devices. While most previous systems used some sort of marker to align virtual content with physical objects, Vazquez et al.~\cite{vazquez2017serendipitous} presented a platform (WordSense) that detects objects in the physical environment and augments them with additional content for language learning including words, sentences, definitions, videos and audio. However, no formal user study was conducted to evaluate the system.

Another example of using AR head-mounted devices for language learning is ARbis Pictus, a system presented by Ibrahim et al.~\cite{ibrahim2018arbis} which labels objects in the user environment with the corresponding vocabulary in the target language. They compared this system to a conventional flashcard-based system and found that AR was more effective and enjoyable and participants could remember words better both shortly after the experiment and four days later. However, the significance of these findings is limited, because the flashcard and AR systems were inherently different. For example, with flashcards the word was shown on the opposite side to the image depicting word meaning thus the image and word were never shown together. This was not the case in the AR condition where word annotations were always visible for all objects in the scene. Therefore it is not clear if AR accounted for the improved performance or the different learning approach. 
%To contribute to this gap in knowledge this paper compares an AR application ruining on a head-mounted device to a a non-AR application that is as similar as possible.   
%To verify the advantage of AR, we will compare our AR application to a non-AR application that is as similar as possible.

In contrast to the presented studies we compare our AR prototype to a non-AR system that is as similar as possible to enable us to measure only the effect of AR without confounding variables like the learning method. To our knowledge, such an experiment has not yet been conducted for AR applications that run on head-mounted devices.

\subsection{Memory Enhancement Techniques}
Research on memory and learning has shown that learning performance and retention depend on different strategies and techniques that can be used to process information in learning~\cite{dunlosky2013}. ``Mnemonic'' is one such technique where the memory capabilities are enhanced by connecting new information to prior knowledge through the use of visuals and/or acoustic cues~\cite{putnam2015,mastropieri1991}. Several researchers experimentally showed that ``Mnemonic'' techniques improve
%for which it has been 
%There, ``Mnemonic'' techniques have been 
%experimentally proven to 
%influential in improving 
memory and recall, especially in the area of language learning~\cite{raugh1975,paivio1981,cohen1980,amiryousefi2011}. 
%Mnemonic is basically an  instructional strategy designed to enhance the memory by connecting new information to prior knowledge through the use of visuals and/or acoustic cues~\cite{putnam2015,mastropieri1991}.

In the field of language learning, mnemonics have mostly been used for  vocabulary learning~\cite{anonthanasap2014mnemonic}. 
%As a combination of visual and acoustic, ``keyword method'' is one of the most effective mnemonic techniques in vocabulary learning~\cite{atkinson1975mnemotechnics}. 
One such mnemonic method is the ``keyword method'' in which learners connect the sound of a word they want to learn to one they already know in either their first language or the target language. 
%In keyword method, students connect the sound of a word they want to learn to one they already know in either their first language or the target language. 
Through this process learners create a mental image  that helps them remember the association~\cite{pressley1982}. For example, the Japanese word for bread is ``sokupan'' which in English sounds like ``Sock + Pan''.  As a result, the learner can imagine a sentence that links a mnemonic keyword with the foreign word. For example, ``sokupan'' can be imagined as frying a sock and putting it on a slice of bread.
%(\autoref{fig:teaser}b).
%The important thing here is that the keyword should clearly relate to the thing being remembered. 

A wide range of existing studies in the broader literature have explored the effectiveness of the keyword method~\cite{atkinson1975mnemotechnics,raugh1975,anonthanasap2014mnemonic,wei2015does}. In this context, comparing the keyword method against other methods in vocabulary learning is one of the most common research designs. There, the keyword method has been compared with learning words in context or learning words with no given strategy. For example, Pressley et al.~\cite{pressley1982mnemonic} found the keyword method to be significantly more effective in learning over the context method. Also, Sagarra and Alba~\cite{sagarra2006key} compared rehearsal, semantic mapping displays and the  keyword method, and found that the keyword method resulted in the best retention. It has also been shown that the keyword method is superior over systematic teaching~\cite{ king1992toward, pressley1982mnemonic}. In 1975, Atkinson and Raugh~\cite{atkinson1975mnemotechnics} found that participants who were given a keyword along with the translation learned more words and also remembered more words after 6 weeks. In the same sense, Raug et al.~\cite{raugh1977teaching} evaluated the use of the keyword method over a long period of 8 to 10 weeks to teach Russian vocabulary and found it to be highly effective. 

Altogether, a significant number of research studies have shown that the keyword method of vocabulary learning is highly effective, yet others showed mixed results~\cite{wei2015does,pearlman1990effectiveness}. For example, a study conducted by Zheng Wei~\cite{wei2015does} found no significant differences between the keyword method, the word-part technique (recognizing part of a word) and the self-strategy.  
%Nevertheless, there is a general agreement in literature that keyword method is effective and that this is the case because it takes advantage of the strength of visual memory. In fact, a number of studies have shown that concrete and easily imagined words (e.g., nouns such as bread) are better remember than more abstract words (e.g., verbs such as walk)~\cite{shepard1967recognition}.

From the perspective of the learning method, VocabulARy builds upon the work of Anonthanasap et al.~\cite{anonthanasap2014mnemonic} in which the authors propose an interactive vocabulary learning system to teach Japanese that automatically creates keywords using phonetic algorithms. There, if the learner selects an image in the system, the phonetically similar words with image representations will gather around the selected image. 
%The word selection is based on the phonetic algorithms. 
Results showed that the keyword-based vocabulary learning system required significantly lower workload than the other compared methods (e.g.\ paper dictionary and static visualisation in a form of an image). 

In summary, our work was inspired by Santos et al. \cite{santos2016augmented}, Vazquez et al. \cite{vazquez2017serendipitous} and Ibrahim et al. \cite{ibrahim2018arbis} who already used AR devices to augment real world objects with annotations for vocabulary learning. 
We combined this approach of providing context with a keyword method 
which has proven to work well in various experiments thus far \cite{atkinson1975mnemotechnics, sagarra2006key,anonthanasap2014mnemonic}. To further advance this learning method we augment keywords with visualisations.
AR provides ideal conditions for that,
because the keyword and its visualisation can be shown in the same
context as the corresponding physical object. 
%To advance aforementioned idea VocabulARy system introduces real-world context (e.g. in kitchen scenario we learn words used in a kitchen) and
%to optimise benefits of keywords
However, in contrast to existing visualisations of keyword approaches we make a careful selection of keywords so that they not only sound similar%(e.g. a Japanese word for bread is ``sokupan'' and sounds similar to keyword ``Sock+Pan'')
, but can also be visualised with an animation in a meaningful way. For example, a Japanese word for bread is ``sokupan'' and sounds similar to keyword ``Sock+Pan'' which can be visualised with an animation of frying a sock in a pan and putting it on a slice of bread. 
We do this to uncover if one can improve vocabulary learning
beyond the influence of the traditional keyword method, by augmenting the
keywords with animated visualisations. According to Shapiro and Waters \cite{shapiro2005investigation} the level of visual imagery of a word enhances vocabulary learning suggesting our approach should work, however no formal evaluation of this exists (~\autoref{tab:relatedwork}).
%and it is also a valid approach to use other visualisations to enhance learning \cite{amiryousefi2011mnemonic}. 
% To the best of our knowledge such animated visualisation of keywords have not been explored this far
%in vocabulary learning context
%. Besides, our system is also the first to combination context based learning with keywords in AR 
%(~\autoref{tab:relatedwork}).
This makes our work both ground breaking and timely.

\section{Augmented Vocabulary Learning / Learning Vocabulary with Visualisations in Real Life Context}

This work presents two prototype systems for vocabulary learning developed on an AR head-mounted-display (HMD) (i.e.\ Microsoft HoloLens 2) and an \rev{$10.5~in$} Android tablet device \rev{(i.e.\ Samsung Galaxy Tab S4)} (\autoref{fig:teaser}). To the best of our knowledge, AR HMD systems for vocabulary learning have not yet been evaluated against comparable non-AR systems (see \autoref{sec:relatedWorkAR}). Both AR and tablet systems combine the keyword method with physical objects. With the AR HMD, our system allows the user to look around a physical room where certain objects are labelled with a button indicating that their translation is available. Upon clicking these buttons with a hand gesture, the English and Japanese word, as well as a keyword with or without visualisation, appears. ~\rev{In addition to the words in both languages and a keyword, an audio of the pronunciation is played. The details of application design and implementation of both prototypes are described in the following sub sections.}

\subsection{Application Design}
\label{sec:application_design}

\rev{In this section we describe key design decisions regarding annotations, animated visualisations and interactions. Careful consideration was given to the selection of annotation and visualisation size. Previous research showed the size of images affects our ability to remember image content during naturalistic exploration~\cite{Shaimaa2022}. However in such exploration individuals are first asked to freely explore an image without any instructions and are then asked about the details observed. As such there is no guarantee that the visual attention is equally distributed across the observed image and as the image gets smaller so does the key information, which makes it easier to miss it. }

\rev{To prevent participants from missing key information we design our application to effectively guide visual attention. The application shows only one word visualisation at a time, which avoids cluttering the scene and overloading participants with too much information. Furthermore, we specifically chose to place AR buttons on the physical surface at close proximity to the object of which word was being memorised. This led users to the appropriate physical location from which AR visualisations are clearly visible as the corresponding annotation and animated visualisation size were appropriated for such viewing. To make the \textsc{non-ar} and \textsc{ar} condition as comparable as possible we made sure that the relative size of annotations and animation was roughly the same in both conditions.}

%\rev{As such all text and visualisation details were clearly visible as long as the system was used in an intended way. We made sure this was the case for both AR and Android prototype. It is important to note that in \textsc{ar} condition the size of annotations and animated visualisations is dependent on the point of view of the observer(e.g. things get smaller as the observer moves away from visualisations). In \textsc{non-ar} condition this is not the case as it is possible to fix the point of view of the observer.}

\rev{Furthermore, in one of the instruction modes, the keyword is also accompanied by its animated visualisation. Such a visualisation consists of a 3D model resembling the keyword and a short animation involving the objects in question. Besides, the user can also listen to the pronunciation of the Japanese word again by clicking a virtual button that is displayed next to the keyword. For example, Figure~\ref{fig:teaser}b shows how the English word ``Postcard" and the corresponding Japanese word ``Hagaki" are displayed. ``Hagaki" sounds like ``Hug + A Key", so it is displayed as a keyword and visualised through ``hugging a key".}

\subsection{Implementation of AR Prototype}
\label{sec:implementation_of_AR}

\begin{figure}[t]
    \includegraphics[width=1\columnwidth]{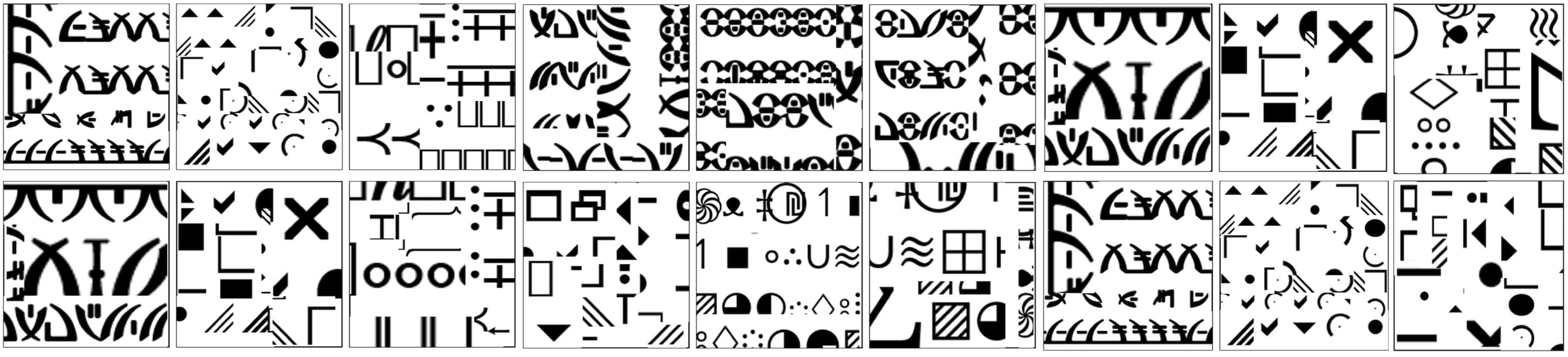}
    \caption{Custom-made image markers.}
    \label{fig:markers}
\end{figure}

\rev{The AR prototype was implemented for Microsoft HoloLens~2\footnote{https://www.microsoft.com/en-us/hololens} using the Unity3D game development engine\footnote{https://unity.com/}. For camera pose tracking, the HoloLens inbuilt tracking system was used. To initialise the positions of augmentations in our application, we used Vuforia\footnote{https://developer.vuforia.com/} and our custom-made image markers (see Figure~\ref{fig:markers})}. \rev{We opted for markers in order to support a reliable and accurate detection of physical objects that correspond to the set of vocabulary. These markers were removed from the scene after initialisation.}

\rev{It would be technically possible to use object recognition techniques to perform object identification and localisation as in~\cite{Castle2009,Salas2013,Salman2018}. Such implementation could support arbitrary environments without prior preparation, which would enable wide implementation of the system.} However, this was not the scope of this work, as we focus on the effect of learning vocabulary using AR and visualisations.

To interact with the virtual contents, we use HoloLens’ built-in hand-tracking and gesture inputs, \rev{which allow the user to interact with virtual content by moving the hands or fingers to content’s corresponding positions. More specifically we chose to use virtual buttons placed on top of planar physical surfaces such as a table or a wall. In this way, touching a surface acted as a tangible feedback making the button press more realistic.}

%Upon identifying an object through the marker, a virtual button is augmented next to the physical object, which the user can press to activate the object, revealing the English and Japanese words. In addition to the words in both languages, the application also displays a keyword that can help the learner to remember the Japanese word better and an audio of the pronunciation is played. In one of the instruction modes, the keyword is also accompanied by its animated visualisation. \rev{Such a visualization consists of a 3D model resembling the keyword and a short animation involving those objects}. Besides, the user can also listen to the pronunciation of the Japanese word again by clicking a virtual button that is displayed next to the keyword. For example, Figure~\ref{fig:teaser}b shows how the English word ``Postcard" and the corresponding Japanese word ``Hagaki" are displayed. ``Hagaki" sounds like ``Hug + A Key", so it is displayed as a keyword and visualised through ``hugging a key".
 
% \begin{figure}[t]
% %\begin{figure}[htb!]
% 	\centering % avoid the use of \begin{center}...\end{center} and use \centering instead (more compact)
% 	\includegraphics[width=1\columnwidth]{figures/ARuser.jpg}
% 	\caption{The office environment in TABLET: The Japanese word for "Postcard" is "Hagaki". The keyword "Hug + A Key" which sounds similar to "Hagaki" helps to remember the word. It is supported by the visualisation of hugging a key. \hl{need to add a better image}}
% 	\label{fig:ARvocabulary}
% \end{figure}

\subsection{Implementation of Android Prototype}
The Android version of the prototype was also implemented using the Unity 3D development environment, but deployed on a Samsung Galaxy Tab S4~\footnote{https://www.samsung.com/global/galaxy/galaxy-tab-s4/}
% \hl{specify the type of tablet} 
tablet device. Its functionality is similar to the AR prototype.
However, instead of seeing the real world environment, an image of an environment is displayed on the screen. In our prototype this was either a kitchen or an office environment. As in the AR prototype certain objects are accompanied with a button. If the user touches the button, the corresponding English and Japanese words, keywords, and animated visualisations appear (Figure~\ref{fig:teaser}c). Visualisations only appear in one instruction mode. \rev{As the size of all objects was adapted to be clearly visible on the screen and to ensure that the relative size of annotations and animation was roughly the same in both conditions (see \autoref{sec:application_design}), we did not provide a feature to zoom into the scene.}

\rev{Compared to the AR implementation, the tablet application can be used anywhere as it can also show scenes that are not related to the real-world environment around the learner. Because all kinds of virtual scenes can be presented, the tablet allows a more flexible use, such as learning words related to a forest while sitting in the living room.}

% \begin{figure}[t]
% %\begin{figure}[htb!]
% 	\centering % avoid the use of \begin{center}...\end{center} and use \centering instead (more compact)
% 	\includegraphics[width=1\columnwidth]{figures/Tab.jpeg}
% 	\caption{The same office environment is displayed as an image on a Tablet with similar functionalities. \hl{need to add a better image}}
% 	\label{fig:Androidvocabulary}
% \end{figure}

\subsection{Generating Keywords}
For generating the keywords, we conducted a small informal survey with 7 participants. They were presented with 28 Japanese-English word pairs and were asked to come up with English words sounding similar to the Japanese words.
At the end, we selected 20 words for the study for which the participants could come up with good keywords.
As already mentioned, the process of finding keywords could also be automatized \cite{anonthanasap2014mnemonic}, but for the scope of this study, this was not needed.

%%%%%%%%%%%%%%%%%%%%%%%%%%%%%%%%%%%%%%%%%%%%%%%%%%%%%%%%%%%%%%%%%%%%%%%%%%%%%%%%%%%%%%%%%%%%%%%%%%%%%%
%%%%%%%%%%%%%%%%%%%%%%%%%%%%%%%%%%%%%%%%%%%%%%%%%%%%%%%%%%%%%%%%%%%%%%%%%%%%%%%%%%%%%%%%%%%%%%%%%%%%%%
\section{Research Method}
This section describes the study conditions, study design, participants’ sampling, study procedure, data collection instruments, and analysis.

\subsection{Study Conditions}
\label{sec:study_conditions}
\begin{figure}[t]
    \includegraphics[width=1\columnwidth]{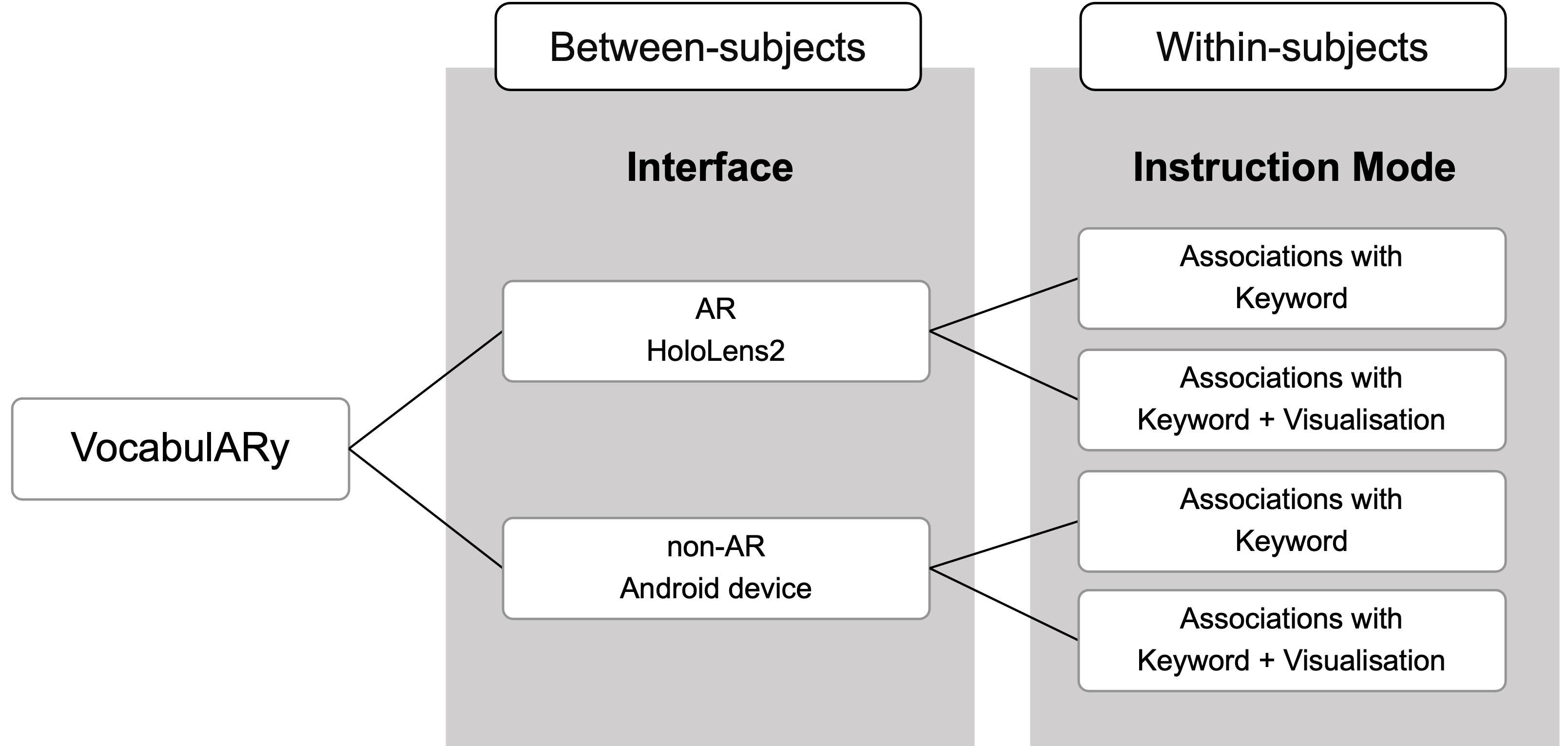}
    \caption{Study design and conditions.}
    \label{fig:design}
\end{figure}

We designed four study conditions based on two distinct vocabulary learning scenarios. The first scenario displays ten physical objects related to a kitchen environment, while the second shows an office environment with ten relevant physical objects. In each of these scenarios a different 
%type of 
\textsc{instruction mode} is provided. This is either a \textsc{keyword} instruction mode or a \textsc{keyword + visualisation} instruction mode. In the \textsc{keyword} condition, only the written keywords are displayed to support the participants in remembering the word. In the \textsc{keyword + visualisation} condition, an animated 3D visualisation of the keyword is displayed in addition to the written keyword. %  a written keyword only or a keyword in combination with it's visualisation. 
These variations are presented to the participants on two different \textsc{Interfaces}, one in \textsc{ar} on a HoloLens2 and one in \textsc{non-ar} on an Android tablet. These study conditions are illustrated in Figure~\ref{fig:design}. 

\subsection{Study Design}
Our study design has two independent variables: \textsc{instruction mode}  which is \textsc{keyword} or \textsc{keyword + visualisation} and  \textsc{interface} which is either \textsc{ar} or \textsc{non-ar}.
%and  
%In the \textsc{ar} condition, participants wear the HoloLens2 and the vocabulary is displayed next to the corresponding physical objects. In the \textsc{non-ar} condition, participants see an image of the physical environment and vocabulary is displayed within that image.
%Repetition of text above
%The second independent variable is \textsc{instruction mode} which is either \textsc{keyword} or \textsc{keyword + visualisation}. In the \textsc{keyword} condition, only the written keywords are displayed to support the participants in remembering the word. In the \textsc{keyword + visualisation} condition, a 3D visualisation of the keyword is displayed in addition to the written keyword.
We used a 2 × 2  mixed design (see Figure~\ref{fig:design}) 
%to evaluate the study conditions, 
because a within-subjects design would make the study, which is mentally demanding, too long (i.e.\ approximately two hours). 
%have resulted in a study duration of approximately two hours. 
We believe that such a long duration could intensify the fatigue and hinder the performance of the participants. Furthermore, running all 4 conditions in a within-subject design would require 2 additional learning scenarios making it more difficult to counterbalance for scenario effects. Reducing the study length by running the study in several sessions also introduces other biases and practical issues.
Therefore, the \textsc{instruction mode} was evaluated as a within-subjects variable and the \textsc{interface} as a between-subjects variable. This means each participant was either using the \textsc{ar}-prototype or the \textsc{non-ar}-prototype, but all participants experienced the \textsc{keyword} and the \textsc{keyword + visualisation} conditions.

To avoid the ``order effects'', which may have an influence on participants' performance due to the order in which the conditions are presented \cite{schuman1981context}, the order of the \textsc{instruction mode} 
%(\textsc{keyword} and \textsc{keyword + visualisation})
as well as the order of the learning scenarios (the kitchen and the office environments) was counter balanced. Special care was given to counterbalance the learning scenario across all independent variables.

\subsection{Participants}
The study was completed by 32 participants, all voluntarily recruited from our university.
%the Faculty of Mathematics, Natural Sciences and Information Technologies, University of Primorska, Slovenia 
None of the participants had any prior knowledge of the Japanese language (identified via a short competency test questionnaire). The between subject sample comprised of 16 participants for the \textsc{ar} condition (10 females) and 16 participants for the \textsc{non-ar} condition (7 females). All the participants were between the age of 19 to 30 years, with the mean of $\overline{x} = 21.6$ and ${SD} = 2.1$. 
%The assignment across all 4 conditions was done randomly. %and randomly assigned to one of the two conditions~(\textsc{ar} or \textsc{non-ar}).

\rev{All our participants were computer science undergraduate and graduate students. No student had previous experience with AR HMDs. The mean age for the \textsc{ar} group was $\overline{x} = 22.13$ (${SD} = 2.68$), and for \textsc{non-ar} group $\overline{x} = 21.13$ (${SD} = 1.26$). The percentage of females in the  \textsc{ar} group was $\overline{x} = 62.5\%$, and in the \textsc{non-ar} group $\overline{x} = 43.75\%$. The groups were comparable. }

\subsection{Procedure}
% First participants signed the consent form and read the PIS \hl{what is pis?}.
% Then they got a short introduction to the study. \hl{what exactly was told?}
% After that they filled out the QCM questionnaire to assess their current motivation.
% Then they used a demo application for 5 minutes to become familiar with the system.
% Before starting with the actual task, a GSR sensor was attached to the participants wrist.
% Then the participant had up to 15 minutes to learn 10 words from the first scenario. Participants could stop earlier if they thought they know all the words already.
% After a one minute break, participants filled out the cognitive load and short-term retention questionnaire.
% Then participants had another 5 minute break, before starting to learn the 10 words from the second scenario for which they also had up to 15 minutes. 
% Again, after a one minute break, they filled out the cognitive load and short-term retention questionnaire.
% After that they answered the user experience and system usability questionnaires as well as the post questionnaire. \hl{what is post questionnaire?}
\rev{On arrival participants were first randomly assigned to one of the two groups (\textsc{ar} or \textsc{non-ar}). Next, we randomly selected which instruction mode would be used first (\textsc{keyword} or \textsc{keyword + visualisation}). Finally, the learning scenario was also randomly chosen (kitchen or office environment). All randomisations were counterbalanced.} 

After being assigned to a particular condition, participants were given a consent form to sign, together with the Participant Information Sheet (PIS) outlining the entire research process in simple language. After briefly explaining the vocabulary learning task with the two learning scenarios, they were asked to fill in the Questionnaire on Current Motivation (QCM)~\cite{rheinberg2001qcm}. 

Before starting the actual task they completed a five-minute training session on a separate demo application to understand the VocabulARy interface. % and the interactions of the system. 
%A galvanic skin response (GSR) sensor was then attached to the participants' wrist. The data from these sensors are out of scope of this paper and will be published in a separate publication. 
Participants were then given up to 15 minutes to complete the first language learning scenario with 10 words. After finishing the first scenario, they filled in the NASA Task Load Index (NASA-TLX) questionnaire~\cite{hart1988development} and, then answered a post-test questionnaire developed by the researchers to assess their immediate recall performance. After taking a 5 minutes break, they were again given up to 15 minutes to complete the second language learning scenario with 10 words. Subsequent to the second scenario, they filled in the same NASA-TLX and the immediate recall questionnaires.

After finishing the experiment, participants were given another two standard questionnaires -- a system usability (SUS)~\cite{lewis2009factor} and a user experience questionnaire (UEQ)~\cite{schrepp2017design}. At the end, participants filled in a short post-questionnaire with demographic questions, questions about previous experience with AR technology and vision problems they might have. The entire experiment took 45 to 60 minutes.

One week later, participants were again requested to answer the same post-test questionnaire developed by the researchers to assess their delayed recall performance as undertaken in prior work \cite{hautasaari2019vocabura}. 
% The overall procedure of the study is shown in Table~\ref{tab:studyProcedure}.

\subsection{Data Collection}
\label{subsec:data_collection}
In order to measure the task completion time,  the time stamp data (start time and end time) were logged by the system. 
%The GSR data was also captured in all conditions (as mentioned, these will be published in our later work).
To measure the motivation, the short form of Questionnaire on Current Motivation (QCM) with 12 items/questions~\cite{rheinberg2001fam,freund2011measure} was used. Anxiety, challenge, interest, and probability of success were measured on a five-point Likert scale, with the labels ``strongly disagree'' at 1 and ``strongly agree'' at 5. Rather than aiming for constructing sub-dimensions (i.e., anxiety, challenge, interest, and probability of success), we used the mean score of the 12 items as an indicator for the overall motivation. 

The NASA Task Load Index (NASATLX)~\cite{hart2006nasa,nasa2006tlx} was used to measure participants’ subjective level of workload/mental effort. Participants rated five of its six dimensions (mental demand, physical demand, temporal demand, effort and frustration) on a 20-point scale ranging from 0 (very low) to 20 (very high). The endpoints of the sixth dimension (own performance) were success and failure. Finally, the overall workload/mental effort was calculated across these six dimensions.

%The items for the retention questionnaire (both short-term and long-term)-- which was developed by the researchers so as to determine participants’ performances-- were prepared based on the Japanese vocabulary they practised during the learning task (e.g., “What is the Japanese word for faucet? What is the Japanese word for paper?”).
In the retention questionnaires, participants were asked for the Japanese translations of the vocabulary they learned. This was undertaken immediately after interacting with the prototype (Immediate Recall) and after one week (Delayed Recall).

\rev{Learning efficiency was determined based on the ratio of performance to the difficulty of the learning task as proposed in~\cite{paas1993efficiency}. The performance of each study condition was based on the recall scores participants obtained after completing the task. The difficulty of the task was based on the mental effort they invested during the learning phase (see \autoref{subsec:ME}). Performance and task difficulty data were then standardised using Formula~\ref{eq:zscore} where $z$ = Z-score,
$r$ = Raw data score, $M$ = Population mean, and $SD$ = Standard deviation.}

\rev{\begin{equation} \label{eq:zscore}
z = \frac{r-M}{SD}
\end{equation}
\\
}

% \rev{ We also measured the learning efficiency for each study condition using Formula~\ref{eq:2D}~\cite{paas1993efficiency,halabi2006applying,clark2011efficiency}. We measured the performance of each study condition based on the recall scores participants obtained after completing the task. We estimated the difficulty of the task based on the mental effort they invested during the task (see subsection \ref{subsec:ME}) and calculated the learning efficiency for each of the four performance conditions.}
% : 
% (a) \textsc{instruction mode} with \textsc{keyword + visualisation} in \textsc{ar}, (b) \textsc{instruction mode} with \textsc{keyword} in \textsc{ar}, (c) \textsc{instruction mode} with \textsc{keyword + visualisation} in \textsc{non-ar} and (d) \textsc{instruction mode} with \textsc{keyword} in \textsc{non-ar} using Formula~\ref{eq:2D}. 

\rev{Next, the learning efficiency $(E)$ was assessed for each of the four study conditions (\autoref{sec:study_conditions}) using  Formula~\ref{eq:2D}~\cite{paas1993efficiency,halabi2006applying,clark2011efficiency} where $E$ = Learning efficiency, $z_P$ = Average performance in Z-scores, and $z_M$ = Average task difficulty in Z-scores. This was done for both immediate recall performance (immediately after participants had completed the task) and delayed recall performance (a week after participants had completed the task). Note that square root of 2 in this formula comes from the general formula for the calculation of distance from a point, $p(x, y)$, to a line, $ax+by+c=0$.}

% The immediate recall and delayed recall learning efficiency results across the study conditions are depicted in Figure~\ref{fig:set4}.}

\rev{\begin{equation} \label{eq:2D}
E = \frac{z_P-z_M}{\sqrt{2}}
\end{equation}
}

To measure the usability of the system, we used the System Usability Scale (SUS), a ten question questionnaire originally created by Brooke, 1996~\cite{brooke1996sus}, on a five-point Likert scale, ranging from ``Strongly'' agree at 1 to ``Strongly disagree'' at 5. For measuring the user experience we used the short version of the User Experience Questionnaire (UEQ-S)~\cite{schrepp2017design,mar2019ueq} with eight items/questions, reported on a 7-point Likert scale. The first four represent pragmatic qualities (Perspicuity, Efficiency and Dependability) and the last four hedonic qualities (Stimulation and Novelty)~\cite{mar2019ueq}.

\subsection{Data Analysis}
The analysis was completed in R studio. Each data set collected in the study was first checked \rev{for mixed ANOVA assumptions. The normality assumption was checked using the Shapiro–Wilk normality test~\cite{sha1965bio}.
Most of the data sets were normally distributed with some of them only approximately normally distributed. 
% however, as some of the data sets were only approximately normally distributed, we used robust statistical methods implemented in ``WRS2'' R package to conduct the analysis, which is a standard practice in such cases. 
The homogeneity of variance assumption of the between-subject factor (\textsc{interface}) was checked using the Levene’s test~\cite{schultz1985} that confirmed homogeneity of variances for each variable ($p > 0.05$). Finally, the homogeneity of covariances of the between-subject factor (\textsc{interface}) was evaluated using the Box’s M-test of equality of covariance matrices. The test showed homogeneity of covariances for each variable ($p > 0.001$). Considering the fact that some of the data sets were only approximately normally distributed, we used robust statistical methods implemented in the ``WRS2'' R package to conduct the analysis~\cite{mair2020robust}, which is a standard practice in such cases.}

%As most of the data-sets were normally distributed whilst some of the data-sets were approximately normally distributed, we used robust statistical methods implemented in ``WRS2'' R package to conduct the analysis.}
In all statistical analyses we used a significance level $p-value > 0.05$ and a restrictive confidence interval (CI) of 95\%. 
%that has implemented various robust statistical methods (i.e., robust ANOVA (including one-way ANOVA, test, between-within subject ANOVA test and etc.), robust t-tests (independent and dependent samples), robust correlation, and nonparametric ANCOVA models based on robust location measures)~\cite{mair2020robust}. 
For immediate recall, delayed recall, mental effort, task completion time and learning efficiency, the statistical significance was examined using a \rev{robust} two-way mixed ANOVA on the 20\% trimmed means--``bwtrim''~\cite{mair2020robust}. 
%The main between-subjects effect (group comparisons), the main within-subjects effect (e.g., due to repeated measurements), and the interaction effect, were computed using ``sppba'', ``sppbb'', and ``sppbi'' functions respectively~\cite{mair2020robust}. 

\rev{For motivation, system usability and user experience, the statistical significance was examined using a Mann–Whitney U test~\cite{sullivan2013}}. 
%The resulting $p < 0.05$ are reported as statistically significant. 
% All plots use a 1.5xIQR (interquartile range) rule and Tukey’s fences~\cite{lisa2016int} for whiskers and identified outliers. 
Asterisk notation is used in tables to visualise statistical significance (ns: $p > 0.05$, *: $p < 0.05$, **: $p < 0.01$, and ***: $p < 0.001$). 

\rev{To assess the reliability of motivation and mental effort questionnaires, we performed a Cronbach’s alpha test. Estimated reliability for each questionnaire (motivation Cronbach’s $\alpha = 0.77 $ and mental effort $\alpha = 0.85 $) is acceptable for research purposes~\cite{ary2018intro}. To measure the reliability of retention questionnaires, we conducted a Kuder-Richardson 20 test~\cite{kuder1937theory}. The $KR = 0.61 > 0.5$ value indicates that the reliability of the retention questionnaire is also acceptable.}

We also conducted a power analysis to check and validate the results and findings of the study. We calculated the effect size (Cohen's $d$) for each data set collected~\cite{cohen1988stat}, selected the minimum effect size (Cohen's $d = 0.69$) and estimated the statistical power ($1-\beta$) of data to check whether the type II error probability ($\beta$) is within an acceptable range for a given sample size ($n=16$ per group) and a significance level ($\alpha = 0.05$). The estimated power value 0.96 shows that with the given sample size, we can have more than a 90\% chance that we correctly reject the null hypothesis with a significance level of 0.05.

%%%%%%%%%%%%%%%%%%%%%%%%%%%%%%%%%%%%%%%%%%%%%%%%%%%%%%%%%%%%%%%%%%%%%%%%%%%%%%%%%%%%%%%%%%%%%%%%%%%%%%
%%%%%%%%%%%%%%%%%%%%%%%%%%%%%%%%%%%%%%%%%%%%%%%%%%%%%%%%%%%%%%%%%%%%%%%%%%%%%%%%%%%%%%%%%%%%%%%%%%%%%%
\begin{figure*}[ht]
%\begin{figure}[htb!]
	\centering 
	\includegraphics[width=2\columnwidth]{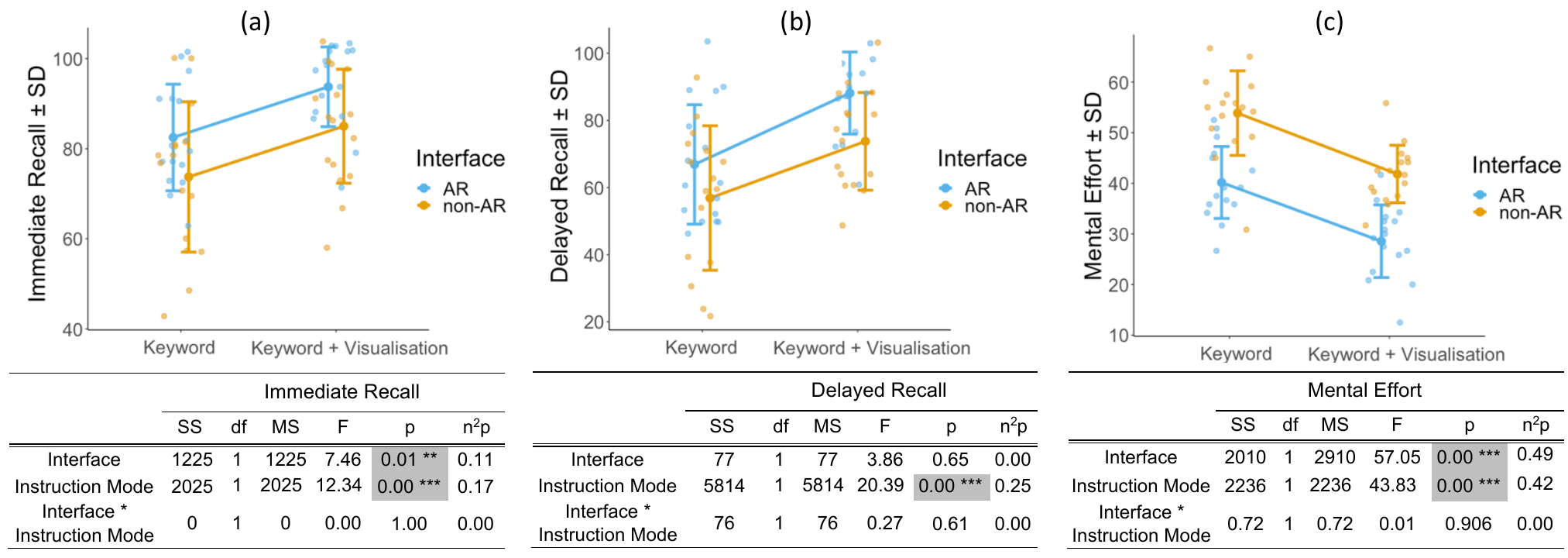}
 	\caption{Means with standard deviation and ANOVA results for: (a) immediate recall performance in percentage of correctly remembered words; (b) delayed recall performance in percentage of correctly remembered words; (c) Mental effort.}
%  	\hl{update this figure}}
	\label{fig:set1}
\end{figure*}

\section{Results}
\label{sec:results}
% \hl{explain how we evaluated data, repeated measures ANOVA, log-transform, ...}

The results and analysis are based on the 32 participants who had completed all the facets of the study, i.e., the motivation questionnaire, the basic training, the vocabulary learning task and the post-questionnaires include mental effort, immediate recall and delayed recall tests. Participants had not undertaken any extra study for the delayed recall test. 

\rev{We conducted a statistical analysis including gender as a between-subject factor (2 (\textsc{gender}) x 2 (\textsc{interface}) x 2 (\textsc{instruction mode}) mixed design) in order to exclude possible gender-based differences. The results did not indicate any statistical significant effect of \textsc{gender} on any dependant variable: immediate and delayed recall, mental effort, task completion time and, immediate and delayed learning efficiency.} The results related to these variables are presented in the following subsections.

\subsection{Immediate Recall}
The mean values of immediate recall performance and the ANOVA results across study conditions, i.e., the \textsc{interface} (\textsc{ar} and \textsc{non-ar}) and, the \textsc{instruction mode} (\textsc{keyword} and \textsc{keyword+visualisation}) are shown in Figure~\ref{fig:set1}a. 
% The data summarised in Figure~\ref{fig:STR} is analysed using a between-within subjects ANOVA on the 20\% trimmed means~\cite{mair2020robust}. These results are presented in Table~\ref{tab:resultsSTR+LTR}.

% \begin{figure}[t]
% %\begin{figure}[htb!]
% 	\centering % avoid the use of \begin{center}...\end{center} and use \centering instead (more compact)
% 	\includegraphics[width=1\columnwidth]{Plots/STR.png}
%  	\caption{Mean short-term retention in percentage of correctly remembered words with standard deviation as error bar. }
% %  	\hl{update this figure}}
% 	\label{fig:STR}
% \end{figure}

A significant main effect of \textsc{interface} on immediate recall performance could be detected \rev{($F(1,60) = 7.46$, $p < 0.05$, $n^{2}p = 0.11$)}. Here, participants' immediate recall scores were significantly better in \textsc{ar} condition ($\overline{x} = 88.13\%$, $SD = 10.34$) compared to the \textsc{non-ar} condition ($\overline{x} = 79.38\%$, $SD = 14.67$).
Also, a significant main effect of \textsc{instruction mode} on immediate recall performance could be detected \rev{($F(1,60) = 12.34$, $p < 0.001$, $n^{2}p = 0.17$)}. Results indicated that participants' immediate recall scores in \textsc{keyword + visualisation} ($\overline{x} = 89.38\%$, $SD = 10.75$) were significantly better than in \textsc{keyword} ($\overline{x} = 78.13\%$, $SD = 14.26$). No significant interaction effect could be found between \textsc{interface} and \textsc{instruction mode} \rev{($F(1,60) < 0.001$, $p > 0.05$, $n^{2}p < 0.001$)}.

\subsection{Delayed Recall}
The mean values of delayed recall performance and the ANOVA results across all study conditions, i.e., the \textsc{interface} (\textsc{ar} and \textsc{non-ar}) and, the \textsc{instruction mode} (\textsc{keyword} and \textsc{keyword + visualisation}) are shown in Figure~\ref{fig:set1}b.
% The data summarised in Figure~\ref{fig:LTR} is analysed using a between-within subjects ANOVA on the 20\% trimmed means~\cite{mair2020robust}. These results are presented in Table~\ref{tab:resultsSTR+LTR}.

% \begin{figure}[t]
% %\begin{figure}[htb!]
% 	\centering % avoid the use of \begin{center}...\end{center} and use \centering instead (more compact)
% 	\includegraphics[width=1\columnwidth]{Plots/LTR.png}
%  	\caption{Mean Long-term retention in percentage of correctly remembered words with standard deviation as error bar. }
% %  	\hl{update this figure}}
% 	\label{fig:LTR}
% \end{figure}

No significant main effect was found between \textsc{interface} on delayed recall performance \rev{($F(1,60) = 3.86$, $p > 0.05$, $n^{2}p < 0.001$)}. The significance was only marginally missed. In addition, the mean values for the \textsc{ar} condition ($\overline{x} = 77.50\%$, $SD = 18.50$) were higher than for the \textsc{non-ar} condition ($\overline{x} = 65.30\%$, $SD = 20.00$).

A significant main effect of \textsc{instruction mode} on delayed recall could be detected \rev{($F(1,60) = 20.39$, $p < 0.001$, $n^{2}p = 0.25$)}. Results indicate that participants' delayed recall scores in \textsc{keyword + visualisation} ($\overline{x} = 80.88\%$, $SD = 13.60$) were significantly better than in \textsc{keyword} ($\overline{x} = 61.88\%$, $SD = 19.65$). No significant interaction effect could be found between \textsc{interface} and \textsc{instruction mode} \rev{($F(1,60) = 0.27$, $p > 0.05$, $n^{2}p < 0.001$)}.

\subsection{Mental Effort}
\label{subsec:ME}
The mean values of mental effort (measured by NASA-TLX) invested to carry out the learning task and the ANOVA results over the study conditions are illustrated in Figure~\ref{fig:set1}c.
% The data summarised in Figure~\ref{fig:ME} is analysed using a between-within subjects ANOVA on the 20\% trimmed means~\cite{mair2020robust} and results are presented in Table~\ref{tab:resultsME}.

% \begin{figure}[t]
% %\begin{figure}[htb!]
% 	\centering % avoid the use of \begin{center}...\end{center} and use \centering instead (more compact)
% 	\includegraphics[width=1\columnwidth]{Plots/ME.png}
% 	\caption{Mean mental effort invested during the task with standard deviation as error bar.}
% 	\label{fig:ME}
% \end{figure}

A significant main effect of \textsc{interface} on mental effort could be detected \rev{($F(1,60) = 57.05$, $p < 0.001$, $n^{2}p = 0.49$)}, such that the mental effort was significantly lower for \textsc{ar} condition ($\overline{x} = 34.36$, $SD = 7.14$) compared to the \textsc{non-ar} condition ($\overline{x} = 47.85$, $SD = 7.02$). Also, a significant main effect of \textsc{instruction mode} on mental effort could be detected \rev{($F(1,60) = 43.83$, $p < 0.001$, $n^{2}p = 0.42$)}. Here, participants' mental effort in \textsc{keyword + visualisation} ($\overline{x} = 35.19$, $SD = 6.42$) was significantly lower than in \textsc{keyword} ($\overline{x} = 47.01$, $SD = 7.73$). No significant interaction effects was found between \textsc{interface} and \textsc{instruction mode} \rev{($F(1,60) = 0.01$, $p > 0.05$, $n^{2}p < 0.001$).}

%\begin{table}[t]
%\begin{figure}[htb!]
%    \centering % avoid the use of \begin{center}...\end{center} and use \centering instead (more compact)
%    % \tiny
%    \small
%    \setlength{\tabcolsep}{12pt}
%        
%        \begin{tabular}{|>{\centering}m{2.25cm}||>{\centering}m{0.23cm}|>{\centering}m{0.3cm}|c|c|c|}
%            %\multicolumn{11}{c}{\small\bfseries\textbf{???}} \\
%            \hline 
%            &\multicolumn{4}{c|}{Mental Effort}  \\
%            \hline 
%            & d$f_{1}$ & d$f_{2}$ & F & p \\%&  %$\eta^2_p$ \\ 
%            \hline 
%            Interface & $1$   & $17.9$  & $60.16$  &    \cellcolor{lightgray}$0.0~***$ \\%&  ?     \\ 
%            \hline
%            Instructional Support & $1$   & $13.28$  & $66.54$  &    \cellcolor{lightgray}$0.0~***$ \\%&  ?     \\ 
%            \hline
%            Interface * Instructional Support &  $1$  & $13.28$  & $0.69$  &  $0.42$   \\%& ?      \\ 
%            \hline 
%        \end{tabular}
%
%    \caption{Between-within subjects ANOVA on the 20\% trimmed means results for mental effort. d$f_1$ = d$f_{effect}$ and d$f_2$ = d$f_{error}$.}
%    \label{tab:resultsME}
%    \vspace{-0.3cm}
%\end{table}

\begin{figure*}[ht]
%\begin{figure}[htb!]
	\centering
	\includegraphics[width=2\columnwidth]{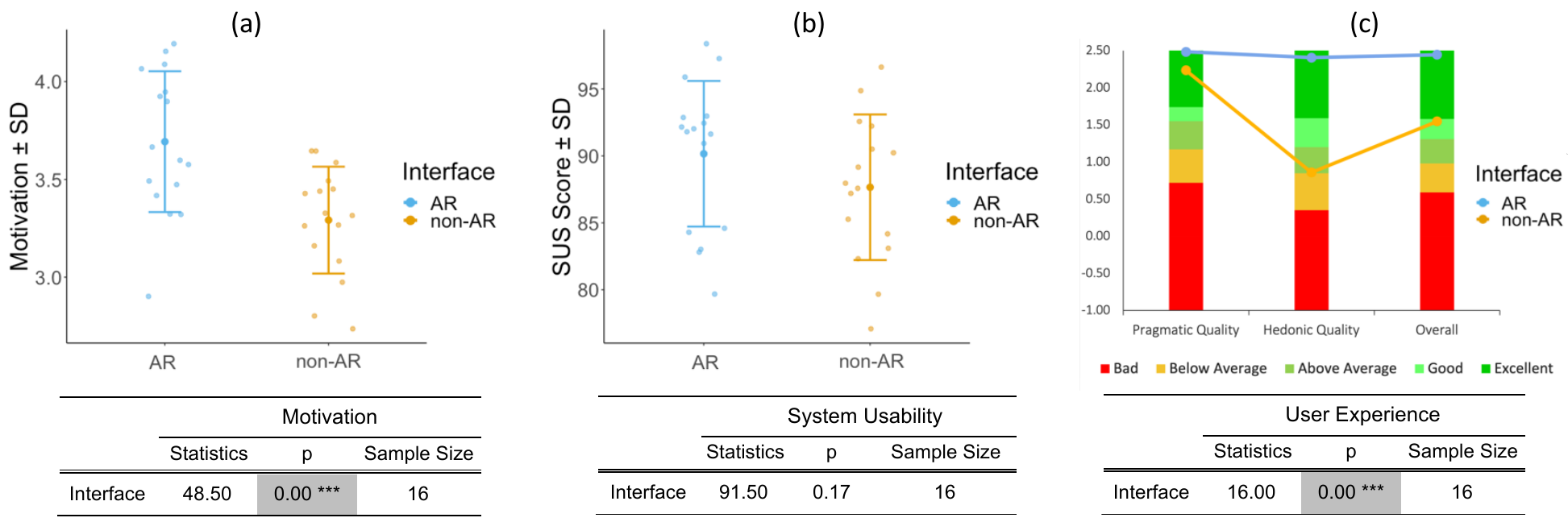}
	\caption{Means with standard deviation for: (a) Motivation before starting the experiment and \rev{Mann–Whitney U test results}; (b) SUS score and \rev{Mann–Whitney U test results}; (c) UEQ factors (pragmatic and hedonic) and all item/question together (overall) with \rev{Mann–Whitney U test results}.}
% 	\hl{update this figure}}
	\label{fig:set2}
\end{figure*}

% \begin{figure}[t]
% %\begin{figure}[htb!]
% 	\centering % avoid the use of \begin{center}...\end{center} and use \centering instead (more compact)
% 	\includegraphics[width=1\columnwidth]{Plots/TCT.png}
% 	\caption{Mean task-completion-time in seconds with standard deviation as error bar.}
% % 	\hl{update this figure}}
% 	\label{fig:TCT}
% \end{figure}

%\begin{table}[t]
%\begin{figure}[htb!]
%    \centering % avoid the use of \begin{center}...\end{center} and use \centering instead (more compact)
%    % \tiny
%    \small
%    \setlength{\tabcolsep}{12pt}
%        
%        \begin{tabular}{|>{\centering}m{2.25cm}||>{\centering}m{0.23cm}|>{\centering}m{0.3cm}|c|c|c|}
%            %\multicolumn{11}{c}{\small\bfseries\textbf{???}} \\
%            \hline 
%            &\multicolumn{4}{c|}{Task-Completion-Time}  \\
%            \hline 
%            & d$f_{1}$ & d$f_{2}$ & F & p \\%&  $\eta^2_p$ \\ 
%            \hline 
%            Interface & $1$   & $17.88$  & $8.87$  &    \cellcolor{lightgray}$0.008~**$ \\%&  ?     \\ 
%            \hline
%            Instructional Support & $1$   & $12.52$  & $32.14$  &    \cellcolor{lightgray}$0.0001~***$ \\%&  ?     \\ 
%            \hline
%            Interface * Instructional Support &  $1$  & $12.52$  & $0.13$  &  $0.73$   \\%& ?      \\ 
%            \hline 
%        \end{tabular}
%
%    \caption{Between-within subjects ANOVA on the 20\% trimmed means results for task completion time. d$f_1$ = d$f_{effect}$ and d$f_2$ = d$f_{error}$.}
%    \label{tab:resultsTCT}
%    \vspace{-0.3cm}
%\end{table}

\subsection{Motivation}
The mean values of motivation between \textsc{Interfaces} (\textsc{ar} and \textsc{non-ar}) is illustrated in Figure~\ref{fig:set2}a. The data summarised in Figure~\ref{fig:set2}a is analysed using a \rev{Mann–Whitney U test}. 
% The results are presented in Figure~\ref{tab:res2}.

% \begin{figure}[t]
% %\begin{figure}[htb!]
% 	\centering % avoid the use of \begin{center}...\end{center} and use \centering instead (more compact)
% 	\includegraphics[width=1\columnwidth]{Plots/MO.png}
% 	\caption{Mean motivation before starting the experiment with standard deviation as error bar. }
% 	\label{fig:MO}
% \end{figure}

A significant effect was found between \textsc{Interfaces} for participants' motivation \rev{($U(N_{AR} = 16, N_{non-AR} = 16) = 48.50$, $p < 0.001$)}. There, the motivation was significantly higher for the \textsc{ar} condition ($\overline{x} = 3.69$, $SD = 0.36$) compared to the \textsc{non-ar} condition ($\overline{x} = 3.29$, $SD = 0.27$).

\subsection{System Usability}
The answers to System Usability Scale (SUS) questions/items are reported on a 5-point Likert scale. The SUS scores are calculated as follows: for each of the odd numbered questions, subtract one from the user response, while for each of the even numbered questions, subtract their response from five and, add up the converted responses for each user and multiply that total by 2.5. This converts possible values to the range of 0 to 100 instead of 0 to 40. These adjustments are kept throughout the rest of the analysis.

% \begin{figure}[t]
% %\begin{figure}[htb!]
% 	\centering % avoid the use of \begin{center}...\end{center} and use \centering instead (more compact)
% 	\includegraphics[width=1\columnwidth]{Plots/SUSScore.png}
% 	\caption{Mean SUS score for \textsc{ar} and \textsc{TAB} \textsc{Interfaces} with standard deviation as error bar.}
% 	\label{fig:SUS}
% \end{figure}

The average SUS scores for \textsc{ar} and \textsc{non-ar} \textsc{Interfaces} are illustrated in Figure~\ref{fig:set2}b. 
% The data summarised in Figure~\ref{fig:set2}b is analysed using a one-way ANOVA test. 
% The results are presented in Figure~\ref{fig:res2}. 
A Mann–Whitney U test indicated that there was no significant effect between the \textsc{Interfaces} for SUS \rev{($U(N_{AR} = 16, N_{non-AR} = 16) = 91.50$, $p > 0.05$)}. 

%\begin{table}[t]
%\begin{figure}[htb!]
%    \centering % avoid the use of \begin{center}...\end{center} and use \centering instead (more compact)
%    % \tiny
%    \small
%    \setlength{\tabcolsep}{12pt}
%        
%        \begin{tabular}{|>{\centering}m{2.25cm}||>{\centering}m{0.23cm}|>{\centering}m{0.3cm}|c|c|c|}
%            %\multicolumn{11}{c}{\small\bfseries\textbf{???}} \\
%            \hline 
%            &\multicolumn{4}{c|}{Motivation}  \\
%            \hline 
%            & d$f_{1}$ & d$f_{2}$ & F & p \\%&  $\eta^2_p$ \\ 
%            \hline 
%            Interface & $1$   & $28$  & $12.6$  &    \cellcolor{lightgray}$0.001~**$ \\%&  ?     \\ 
%            \hline 
%        \end{tabular}
%
%    \caption{One-way ANOVA results for motivation. d$f_1$ = d$f_{effect}$ and d$f_2$ = d$f_{error}$.}
%    \label{tab:resultsMO}
%    \vspace{-0.3cm}
%\end{table}

\subsection{User Experience}
\rev{The UEQ-s questionnaire provides a benchmark to compare user experience between different systems~\cite{hinderks2018}. It measures pragmatic qualities of a system (including efficiency, perspicuity and dependability) and hedonic qualities (including stimulation and novelty). The overall value was calculated from all 5 UEQ scale means. We adapted the standard method as suggested in~\cite{schrepp2017design,mar2019ueq} for calculating the scale means for each factor individually (efficiency, perspicuity, dependability, stimulation and novelty) and to obtain values for pragmatic quality, hedonic quality and overall user experience for both \textsc{ar} and \textsc{non-ar} systems.}

In the \textsc{AR} condition the pragmatic ($\overline{x} = 2.48$) and hedonic  ($\overline{x} = 2.41$) qualities, as well as overall user experience ($\overline{x} = 2.45$) were all perceived as \rev{excellent} \rev{(benchmarks for an excellent score: pragmatic $>$ 1.73, hedonic $>$ 1.55, overall $>$ 1.58)}. In the \textsc{non-ar} condition the mean value of pragmatic \rev{quality was perceived as excellent ($\overline{x} = 2.23$), and the} overall experience \rev{as good ($\overline{x} = 1.55$)} \rev{(benchmarks for a good score: pragmatic between 1.55 - 1.73, hedonic between 1.25 - 1.55, overall between 1.4 - 1.58)}. However, the hedonic factor was \rev{perceived as below average ($\overline{x} = 0.88$)} \rev{(benchmarks for below average score: pragmatic between 0.73 - 1.14, hedonic between 0.88 - 1.24, overall between 0.68 - 1.01)}. 

\rev{The overall user experience was analysed using a Mann–Whitney U test. The data is presented in \autoref{fig:set2}c. A significant effect was found between \textsc{Interfaces} ($U(N_{AR} = 16, N_{non-AR} = 16) = 16.00$, $p < 0.001$).}

\subsection{Task Completion Time}
% The results of the study regarding task completion time can be seen in figure \ref{fig:STR}.
% The results of the repeated measures ANOVA can be seen in table \ref{tab:resultsSTR+LTR}.
The mean values of task completion time across study conditions, i.e., the \textsc{interface} (\textsc{ar} and \textsc{non-ar}) and, the \textsc{instruction mode} (\textsc{keyword} and \textsc{keyword + visualisation}) are shown in Figure~\ref{fig:set3}a. The data summarised in Figure~\ref{fig:set3}a is analysed using a between-within subjects ANOVA on the 20\% trimmed means~\cite{mair2020robust}. 
% These results are described in Figure~\ref{fig:set3}a.

A significant main effect of \textsc{interface} on task completion time could be detected \rev{($F(1,60) = 14.06$, $p < 0.001$, $n^{2}p = 0.19$)}. Here, the completion time was significantly lower for \textsc{ar} condition($\overline{x} = 618.57 s$, $SD = 77.92 s$) compared to the \textsc{non-ar} condition ($\overline{x} = 698.41 s$, $SD = 90.59 s$).
Also, a significant main effect of \textsc{instruction mode} on task completion time could be detected \rev{($F(1,60) = 31.19$, $p < 0.001$, $n^{2}p = 0.34$)}, such that \textsc{keyword + visualisation} ($\overline{x} = 599.04 s$, $SD = 74.99 s$) resulted in a significantly lower completion time than \textsc{keyword} ($\overline{x} = 717.95 s$, $SD = 93.52 s$). There was no significant interaction effect found between \textsc{interface} and \textsc{instruction mode} \rev{($F(1,60) = 0.25$, $p > 0.05$, $n^{2}p < 0.001$)}.

\subsection{Learning Efficiency}
% \begin{figure}[t]
% %\begin{figure}[htb!]
% 	\centering % avoid the use of \begin{center}...\end{center} and use \centering instead (more compact)
% 	\includegraphics[width=1\columnwidth]{Plots/STREfficiency.png}
% 	\caption{Average learning efficiency for short-term retention with standard deviation as error bar.}
% 	\label{fig:STRE}
% \end{figure}
\begin{figure*}[ht]
%\begin{figure}[htb!]
	\centering % avoid the use of \begin{center}...\end{center} and use \centering instead (more compact)
	\includegraphics[width=2\columnwidth]{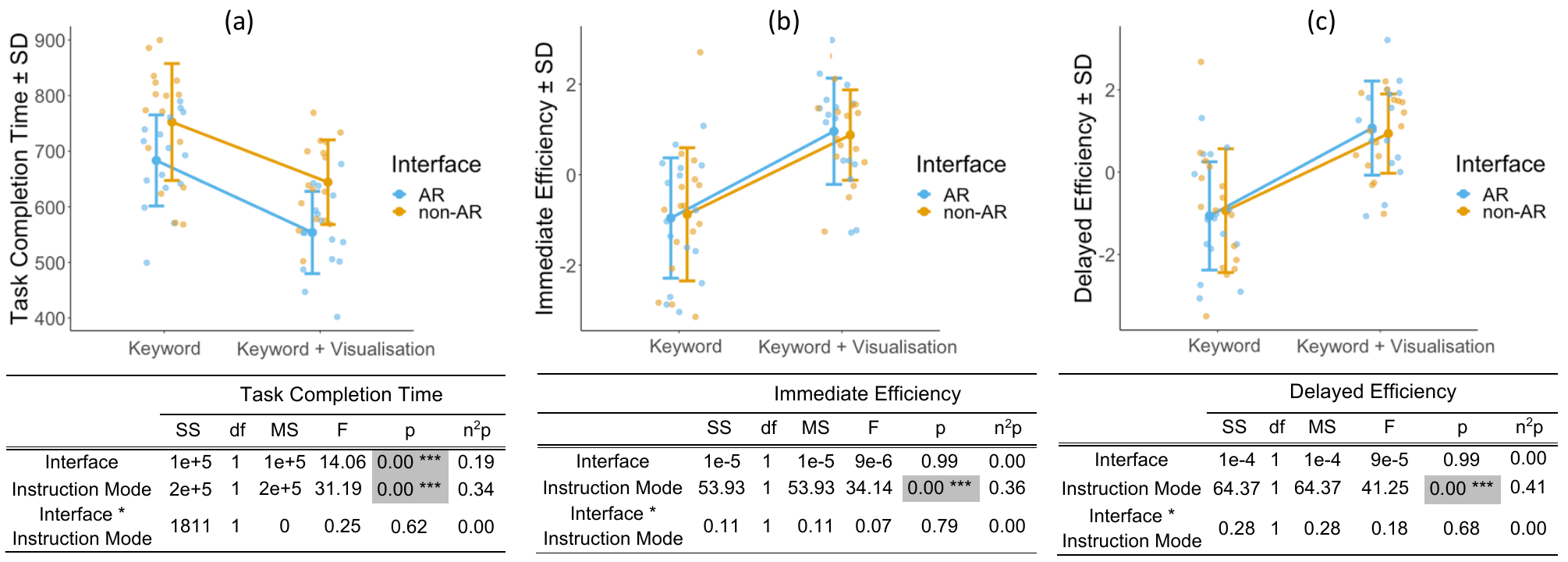}
	\caption{Means with standard deviation and ANOVA results for: (a) Task-completion-time in seconds;  (b-c) immediate recall and delayed recall learning efficiency.}
% 	\hl{update this figure}}
	\label{fig:set3}
\end{figure*}
\begin{figure*}[ht]
%\begin{figure}[htb!]
	\centering % avoid the use of \begin{center}...\end{center} and use \centering instead (more compact)
	\includegraphics[width=2\columnwidth]{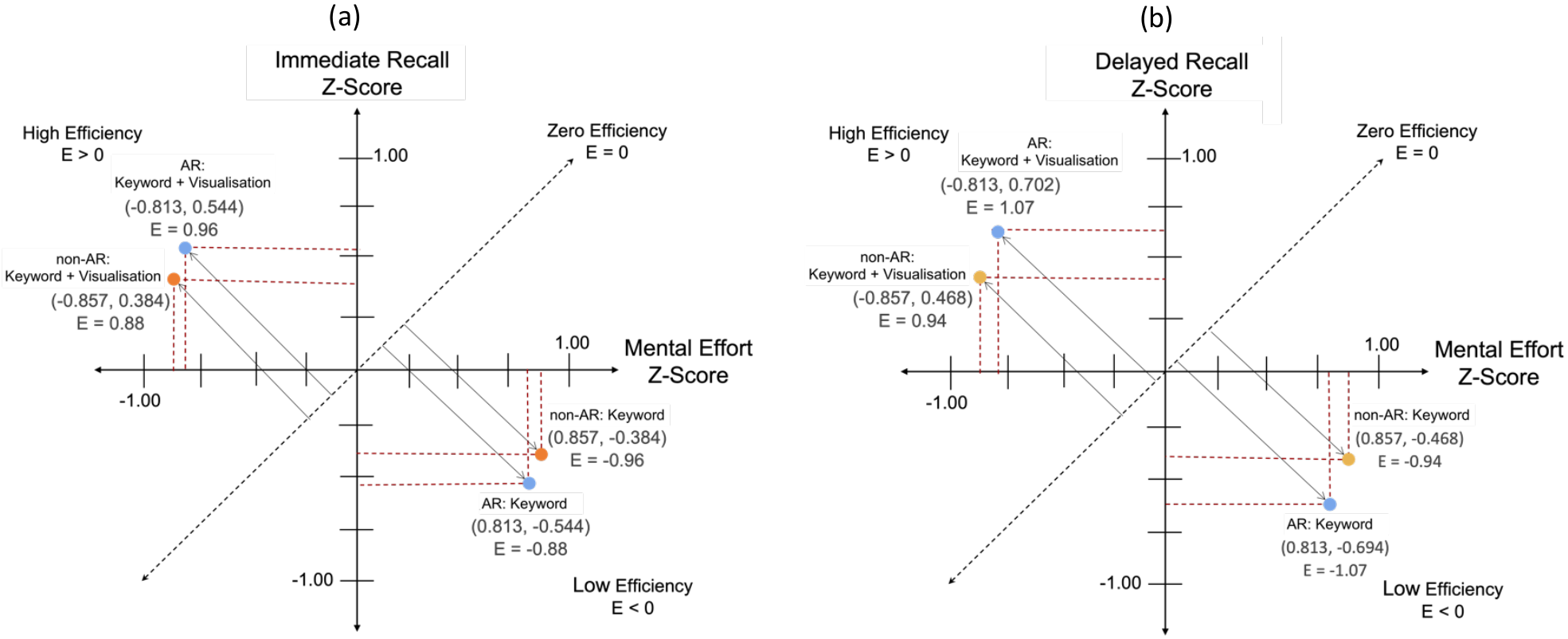}
	\caption{Learning efficiency: (a) immediate recall learning efficiency; (b) delayed recall learning efficiency.}
% 	\hl{update this figure}}
	\label{fig:set4}
\end{figure*}

% \begin{figure}[t]
%     \centering
%         \centering
%         \includegraphics[width=1\columnwidth]{Plots/STREF.png}
%         \caption{Short-term retention learning efficiency.}
%     \label{fig:Efficicency_Short}
% \end{figure}

% \begin{figure}[t]
%     \centering
%         \centering
%         \includegraphics[width=1\columnwidth]{Plots/LTREfficiency.png}
%         \caption{Average learning efficiency for long-term retention with standard deviation as error bar.}
%     \label{fig:LTRE}
% \end{figure}

% \begin{figure}[t]
%     \centering
%         \centering
%         \includegraphics[width=1\columnwidth]{Plots/LTREF.png}
%         \caption{ Long-term retention learning efficiency.}
%     \label{fig:Efficicency_Long}
% \end{figure}

The average learning efficiencies for immediate recall and delayed recall across study conditions
%, i.e., the \textsc{interface} (\textsc{ar} and \textsc{non-ar}) and, the \textsc{instruction mode} (\textsc{keyword} and \textsc{keyword + visualisation})
are shown in Figure~\ref{fig:set3}a and  Figure~\ref{fig:set3}b respectively. \rev{For the definition of learning efficiency refer to~\autoref{subsec:data_collection}}. The data summarised in Figure~\ref{fig:set3}a and Figure~\ref{fig:set3}b are analysed using a between-within subjects ANOVA on the 20\% trimmed means~\cite{mair2020robust}.
% These results are presented in Figure~\ref{fig:set3}.%Table~\ref{tab:resultsSTR+LTR}.

Statistical analysis in Figure~\ref{fig:set3}b and Figure~\ref{fig:set3}c showed no significant effect of \textsc{interface} for participants' learning efficiency for immediate recall \rev{($F(1,60) = 9e-6$, $p > 0.05$, $n^{2}p < 0.001$)} or delayed recall \rev{($F(1,60) = 9e-5$, $p > 0.05$, $n^{2}p < 0.001$)}. A significant main effect of \textsc{instruction mode} on participants' learning efficiency for immediate recall could be detected \rev{($F(1,60) = 34.14$, $p < 0.001$, $n^{2}p = 0.36$)}. There, the learning efficiency was significantly higher in \textsc{keyword + visualisation} support ($\overline{x} = 0.92$, $SD = 0.23$) compared to the \textsc{keyword} ($\overline{x} = -0.92$, $SD = 0.23$). A significant main effect on participants' learning efficiency for delayed recall also could be detected \rev{($F(1,60) = 41.25$, $p < 0.001$, $n^{2}p = 0.41$)}. There, the learning efficiency was significantly higher in \textsc{keyword + visualisation} instruction mode ($\overline{x} = 1.07$, $SD = 1.15$) compared to the \textsc{keyword} ($\overline{x} = -1.07$, $SD = 1.32$) mode. There was no significant interaction effect found between \textsc{interface} and \textsc{instruction mode} for in immediate recall \rev{($F(1,60) = 0.07$, $p > 0.05$, $n^{2}p < 0.001$)} or delayed recall \rev{($F(1,60) = 0.18$, $p > 0.05$, $n^{2}p < 0.001$)}.

\section{Discussion and future directions}
For this study, we developed an AR system called VocabulARy, that supports learning new Japanese words, but can be expanded to support other languages. The system was used to evaluate user experience, system usability, mental effort, motivation and memory recall
%users' preference and recall 
when shown keywords over the objects vs. keywords together with a visualisation of the objects. These were compared in two interfaces: \textsc{ar} (Microsoft HoloLens 2) and a \textsc{non-ar} (Android tablet computer). We used the two interfaces to investigate whether showing keywords and visualisations in context of immediate surrounding compared to the context provided on the virtual scene on the screen results in any performance difference.

\subsection{Usability and User Experience}
The results of the study show that participants evaluated both \textsc{ar} and \textsc{non-ar} prototypes with good usability scores, clearly higher than average (68) and no significant difference between the two could be found. \rev{In addition, during the study we did not observe any readability problems (e.g. none of the users tried to zoom in on the tablet computer in order to make it easier to view presented information and none of the AR HMD users were observed to move very close to augmentations).
This provided a good} basis for further investigation, as we wanted to make the comparison as fair as possible by trying to not influence learning performance with usability issues \rev{as well as by making both conditions as comparable as possible (see \autoref{sec:application_design})}.

It has been shown for example that the unfamiliarity with AR could result in lower performance as has been reported in prior work~\cite{wille2014prolonged}. In addition, the user experience in both conditions has been rated very positively with a higher score for \textsc{ar} regarding the hedonic factors represented by Stimulation and Novelty. This makes sense, as \textsc{ar} is still an exciting and less widely used technology compared to tablet computers for many users.

\rev{A recent study also revealed that the size of images affects our ability to remember image content during naturalistic exploration~\cite{Shaimaa2022}. Although our study did not involve naturalistic exploration, it clearly steered participants’ attention, and we made extra care that both \textsc{ar} and \textsc{non-ar} showed comparable imagery, it would still be interesting to explore if the size of imagery has an effect on the ability to memorise vocabulary words.}

The prolonged use of HMDs in the current form factor can also influence the usability and thus performance of users. %This provides another implication for the design of such systems in longitudinal studies. 
Some studies have already investigated the effects of the HMDs weight, their pressure on the face, latency, image quality and the authenticity of the representation of digital objects~\cite{guo2019mixed, guo2019evaluation, shen2019mental, steinicke2014self}. However, these issues will likely be addressed with future development of HMDs. 

\subsection{AR vs Non-AR}
Our results show that immediate recall (a recall of words right after the study) in the \textsc{ar} system is significantly higher compared to the \textsc{non-ar} system. However, no statistical significance was detected for delayed recall (a recall of the same words a week after the study). Nonetheless, it is important to note that significance was only marginally missed for delayed recall. The results can thus not confirm the outcomes of a previous study conducted by Ibrahim et al.~\cite{ibrahim2018arbis} who report a significantly better performance of the \textsc{ar} system compared to \textsc{flashcards} for both immediate and delayed recall. One of the reasons, and also a major difference between this and our study, might be that in the aforementioned work, the learning methods were not identical in both conditions, which could have placed the AR system at an advantage. For example, with flashcards the word was shown on the opposite side to the image depicting word meaning; thus, the image and word were never shown together. This was not the case in our \textsc{ar} condition where the word annotations were always visible for a selected object in the scene. In comparison, we carefully designed our experiment to minimise any such confounding variable that might influence the results. 

Furthermore, participants expressed a significantly higher level of motivation in the \textsc{ar} condition. This is in line with previous work~\cite{dalim2020using, li2014pilot} and should be considered when interpreting our results since motivation can be an important factor in learning~\cite{chen2009effect} and technology can play a significant role in this~\cite{lin2017study}. What causes higher motivation falls out of the scope of this study; however, one could hypothesise that the novelty of the AR plays an important role. The observations show that participants were excited about testing the AR HMD compared to using a tablet. This introduces a need for a longitudinal study of using AR in vocabulary learning, \rev{as the influence of motivation might decrease with increasing familiarity with the system}. 

Interestingly, the \textsc{ar} condition also outperformed the \textsc{non-ar} condition in task completion time (about 11\% faster). This results show that participants were able to learn all words faster in \textsc{ar} condition compared to the \textsc{non-ar} condition whilst also achieving higher immediate recall scores. This result is also somewhat surprising as \textsc{ar} is at a disadvantage to \textsc{non-ar} for activating objects. That is, target selection in our setup was typically faster on a tablet computer compared to in mid-air tapping on a HMD. Furthermore, a tablet computer also offered instant access to all buttons at the same distance, whereas users need to physically move to activate some of the buttons in AR. \rev{One future direction could involve making users spend the same amount of time in both conditions and explore if this would further improve the performance of \textsc{ar} condition.}

% Finally we also analysed our results from the perspective of learning efficiency where no significant difference could be found between the two interfaces for immediate and delayed recall. The mean values for \textsc{ar} and \textsc{non-ar} are almost identical for both immediate and delayed recall (see \autoref{fig:set3}b and c). %\hl{(add more about what this means.)}

% These finding support our argument that augmented reality can be a valuable addition in vocabulary learning.

%and that mean delayed recall was higher in AR condition for both instruction modes (\textsc{keyword} and \textsc{keyword + visualisation}). 

% regarding immediate recall, mental effort and time needed to study. 

%We predict that the effect of \textsc{interface} drops in learning efficiency if compared to learning performance (immediate and delayed recall) because of increased mental effort users experience in \textsc{ar} interface. WHY??

\subsection{Keyword vs. Keyword + Visualisation}
Regarding our second independent variable \textsc{instruction mode}, our results clearly show that vocabulary learning can be improved beyond the traditional keyword method, by augmenting the keywords with animated 3D visualisations. We found overwhelming evidence for this in all metrics, such as immediate and delayed recall, learning efficiency, mental effort, and even task completion time.
% showing animated visualisations besides the keyword in \textsc{keyword + Visualisation} instruction mode contributed to better immediate and delayed recall.
% outperformed the basic \textsc{keyword} method with respect to immediate recall, \hl{?delayed recall?}, mental effort and time needed to study. In addition, the learning efficiency for immediate recall was significantly higher for \textsc{keyword + visualisation}.
% This suggests that augmenting the keyword with animated 3D visualisations related to the vocabulary can increase learning performance regardless of the interface used. 
This is in line with observations by Shapiro and Waters~\cite{shapiro2005investigation}, who reported that the level of visual imagery of a word enhances vocabulary learning. In this work, we go beyond simple imagery and show that the 3D animated content is potentially an even more effective approach. This opens up another direction for future work involving the necessity for detailed comparison of the effect of different visualisation techniques. % should be explored in more details within future studies. 

As mentioned, our results showed that providing visualisations for keywords reduced the mental effort for vocabulary learning. However, reducing mental effort in learning scenarios can also result in reduced learning outcomes. For example, Salmon showed that the amount of invested mental effort positively correlates with learning efficiency~\cite{salomon1984television}. Knowing this, we could expect a decline in performance of immediate and delayed recall. One reason why this did not happen in our case might be the fact that enough effort was needed in order to complete the task (moving, tapping, remembering). One way to increase the mental effort would be to require users to come up with their own associations for keywords instead of providing predefined keywords as in our study. %participants did not put sufficient amount of effort into visualising keywords by themselves. This could be due to the lack of appropriate skills or the fact that keywords were predefined and not 
Providing predefined keywords might not be in line with user's mental model, thus making it difficult for the user to (mentally) visualise them. This could have made visualisations in our study more important.  However, previous research suggests that users might have difficulties coming up with their own keywords and that predefined keywords lead to better learning outcomes~\cite{atkinson1975mnemotechnics}. Despite, future studies should explore if the difference persists also when personalised keywords are used in learning scenario presented in this study.

%\subsection{Research and Design Implications}

%The prolonged use of HMDs in the current form factor can also influence the usability and thus performance of users. This provides another implication for the design of such systems in longitudinal studies. Some studies have already investigated the effects of the HMDs weight, their pressure on the face, and the 

\subsection{Implications and design recommendations}
The benefits of the keyword method over other learning techniques are well known~\cite{ king1992toward, pressley1982mnemonic}. We have shown that the keyword method can provide even better results by adding animated visualisations that depict the keyword itself. This is an important implication for designing such applications for vocabulary learning. However, this also opens up several questions. For example, do animations of visualisations of keywords contribute to the learning outcome or would visualisation without an animation result in comparable efficiency? 

\rev{One of the most important things to consider in designing such a system is the keywords or words from the language a learner speaks that sound similar to the word being learnt.} In our prototype, we only used a limited set of vocabulary for which we were able to find appropriate keywords and accompanying visualisations.
Finding these keywords and visualisations takes time, which needs to be considered when thinking about applying this method in practice. 
And it is no necessary that every word would have an appropriate keyword. 
Crowd-sourcing could be one approach to tackle this problem. Additionally, approaches for automating the process of finding keywords already exist~\cite{anonthanasap2014mnemonic}. Also, our future direction will involve investigating the effect of asking users to choose their own keywords and visualisations. A system that would use a combination of these approaches could probably satisfy a variety of learning types. 

\rev{Another thing to keep in mind, and we are not aware of any study investigating it, is the fact that the vast number of keywords might be overwhelming for users. One of the unanswered question is thus how many keywords is recommended to provide at one time (in our study only one was shown at the time to direct users).}
%However, as word meaning does not change it could be worth setting this up once and providing it to a very large group of learners. 

\rev{For the purpose of this study we used marker based tracking to initialise the settings in AR. As such, our system was linked to a particular physical space. To enable wider adoption, another space-independent object recognition technique should be used as discussed in \autoref{sec:implementation_of_AR}. Such a system would also need to have a database of objects with the corresponding keywords available upfront so when users look at a physical scene AR visualisations would be fetched on the fly.}

Only two participants used all the available time to learn and all 10 words were correctly remembered in 23\% of immediate and 4\% of delayed recall tests. For immediate recall we might have reached the ceiling effect, and making the task more difficult would highlight even greater changes between the test conditions. This was even more obvious for delayed recall, where only very few users finished the test with no errors. \rev{This could be taken in consideration when building such a system -- during the testing phase the system should try to increase the level of engagement and encourage users to take more time, while and after testing the system should encourage users to rethink about wrong answers.}

%To better understand if participants could learn even more using the proposed vocabulary learning technique, we are planning to conduct a similar experiment, where, within a given time limit, users would try to memorise as many words as possible. 

%While learning in the context of the immediate surroundings of the real environment brings some benefit to the learning efficiency, it might be tiresome for users to use such a system due to the gorilla arm effect of mid-air tapping~\cite{hincapie2014consumed}. The effect states that the prolonged use of arms in front of the user without any support can contribute to arm fatigue and a feeling of heaviness. One approach to tackle this is to study the actual fatigue with consumed endurance metric~\cite{hincapie2014consumed}. Another approach is to study substitute input techniques in our AR prototype to avoid such issues. Some directions have already been proposed in the literature~\cite{hansberger2017dispelling}. 

\subsection{Limitations}

\rev{As explained in \autoref{sec:results} gender did not have a significant effect on the results of the study. However, future work should look into a possible gender bias in more detail with a higher number of participants, as our result on this is not conclusive.}

\rev{Another thing to consider in our study is age bias. However, the age group studied is highly mobile, spending extended periods of time in foreign countries (for example, the EU Erasmus+ programme alone funds more than half a million exchanges yearly~\cite{Erasmus}). As such, this group could benefit from an improved vocabulary learning system. Nevertheless, the results cannot be generalised over the whole population, and expanding the study to other age groups and exploring the effect of age on the proposed learning system is an important future direction.}

%\rev{Our prototype currently supports language learning in a specific environment equipped with markers for initialising the system. To make it suitable for a broader range of users it should support variable environments which can be obtained by implementing object detection to recognize relevant objects, as was mentioned in the implementation section.}

Further, we only used nouns in our prototype. More specifically, all nouns were associated with objects. In fact, a number of studies have shown that concrete terms (e.g., nouns such as bread) are better remembered than abstract terms (e.g., abstract nouns and verbs)~\cite{shepard1967recognition}. The benefits of in-situ learning with AR will therefore be reduced when abstract terms are considered as it becomes difficult to make them relevant to context of users' immediate environment. \rev{ Nevertheless, future studies could focus on exploring the potential of 3D AR animation to make abstract terms visually more accessible.}

% As indicated by Shapiro and Waters \cite{shapiro2005investigation} the level of visual imagery of words strengthens the keyword method. \hl{check if they mean words to learn or keywords - both would apply in our case - and discuss this further}.

%Discuss ceiling effects? Could some participants have learned even more?
As mentioned in the paper, our prototype was only tested for a short time on a limited vocabulary. To further validate our findings, the vocabulary should be expanded and tested over a longer period of time. Especially the effect of higher motivation in \textsc{ar} could wear off as users become more familiar with the system.
%In our prototype, users were presented with previously chosen keywords and their visualisations. Future work should investigate the effect of choosing their own keywords and visualisations. This could be done either by enabling them to insert their own or by giving them a choice of several keywords.
Additionally, we only measured the immediate recall (immediately after participants had completed the task) and a delayed recall (a week after participants had completed the task) of the vocabulary learned. Future work should also consider recall after longer periods of several weeks. This could also be combined with repeating the learning phase in certain intervals, as it is normally done when learning vocabulary.

%Findings:
%\begin{itemize}
%    \item STR: AR better than Tab; visualisation better than Keyword
%    \item LTR: \hl{not yet analyzed}
%    \item Mental Effort: AR lower than Tab; visualisation lower than keyword
%    \item TCT: AR less time than tab; visualizaiton less time than keyword
%    \item Motivation: participants significantly more motivated for AR than Tab
%    \item Learning efficiency STR: no difference between AR and TAB; higher for visualisation than keyword
%    \item Learning efficiency LTR: \hl{not yet analyzed}
%    \item User Experience: pragmatic, hedonic and overal UE are perceived as extremely positive in AR; in non-AR pragmatic and overall are also high
%    \item System Usability: no significant difference between AR and tab
%\end{itemize}

%Other points:
%\begin{itemize}
    % \item ceiling effects? could some participants have learned even more?
    % \item we should definitely expect AR to be better for delayed recall, as found by \cite{hautasaari2019vocabura, santos2016augmented, ibrahim2018arbis}
    %\item did we find differences in gender? (\cite{tabatabaei2011gender} found that women perform better)
    % \item implications
    % \item generalizability
%\end{itemize}

% \section{Limiations}

%%%%%%%%%%%%%%%%%%%%%%%%%%%%%%%%%%%%%%%%%%%%%%%%%%%%%%%%%%%%%%%%%%%%%%%%%%%%%%%%%%%%%%%%%%%%%%%%%%%%%%
%%%%%%%%%%%%%%%%%%%%%%%%%%%%%%%%%%%%%%%%%%%%%%%%%%%%%%%%%%%%%%%%%%%%%%%%%%%%%%%%%%%%%%%%%%%%%%%%%%%%%%
\section{Conclusion}
Learning vocabulary can be enhanced when encountering words in context. This context
can be afforded by the place or activity people are engaged with. 
%Existing learning environments include formal learning, mnemonics,
%flashcards, use of a dictionary or thesaurus, all lead to practice with new words in context. 
For this purpose we developed VocabulARy, a HMD AR system that 
% propose an enhancement
% to the language learning process by providing the user with words and learning tools in context with VocabulARy. VocabulARy 
visually annotates objects in the user’s surroundings, with the corresponding English (first language) and Japanese (second language)
words to enhance the language learning process. In addition to the written and audio description of each word, we also present the
user with a keyword and its animated 3D visualisation to enhance memory retention. 

We evaluated our prototype by comparing it to an alternate AR
system that does not show any additional visualisation of the keyword, and also, we compare it to two non-AR systems on a tablet, one
with and one without visualising the keyword. Our results indicate that \textsc{ar} outperforms the \textsc{non-ar} system regarding short-term retention,
mental effort and task-completion time. Additionally, the visualisation approach scored significantly higher than only showing the written
keyword with respect to immediate and delayed recall, learning efficiency, mental effort and task-completion time. Visualisation of keywords thus proved more efficient compared with the traditional keyword method only and opens new avenues for future improvements in AR enabled vocabulary learning systems.

\section*{Acknowledgement}
We thank Cuauthli Campos for helping us with preparing the video and all the volunteers who participated in the user study.

This research was supported by European Commission through the InnoRenew CoE project (Grant Agreement 739574) under the Horizon2020 Widespread-Teaming program and the Republic of Slovenia (investment funding of the Republic of Slovenia and the European Union of the European Regional Development Fund). We also acknowledge support from the Slovenian research agency ARRS (program no. BI-DE/20-21-002, P1-0383, J1-9186, J1-1715, J5-1796, and J1-1692).

\balance 
%\acknowledgments{
%The authors wish to thank A, B, and C. This work was supported in part by
%a grant from XYZ (\# 12345-67890).}

%\bibliographystyle{abbrv}
\bibliographystyle{abbrv-doi}
\typeout{}
\bibliography{template}
\end{document}